\documentclass[pre,aps]{revtex4}
\usepackage{epsfig}

\begin{document}

\newcommand {\be}[1] {\begin{equation}\label{#1}}
\newcommand {\ee} {\end{equation}}
\newcommand {\ba}[1] {\begin{eqnarray}\label{#1}}
\newcommand {\ea} {\end{eqnarray}}
\newcommand {\Eq}[1] {Eq.~(\ref{#1})}
\newcommand {\Fig}[1] {Fig.~\ref{#1}}
\def \bp {\mbox{\boldmath $\partial$}}
\def \q {{\bf q}}
\def \F2 {FPL${}^2$ }
\def \Rs {\sf I\hskip-1.5pt R}
\def \Zs {\mbox{\sf Z\hskip-5pt Z}}
\def \Cs {\rm C\!\!\!I\:}
\def \rb {\rm b}
\def \rg {\rm g}
\def \OMIT #1{}
\def \rem #1 {{\it #1}}

\title {Conformal field theory of the Flory model of polymer melting}

\author{Jesper Lykke Jacobsen}
\affiliation{Laboratoire de Physique Th\'eorique et Mod\`eles Statistiques\\
         Universit\'e Paris-Sud \\
         B\^atiment 100, F-91405 Orsay, FRANCE \\
         JACOBSEN@IPNO.IN2P3.FR}
\author{Jan\'{e} Kondev}
\affiliation{Physics Department, MS057,  Brandeis University \\
        Waltham, MA 02454, USA \\
        KONDEV@BRANDEIS.EDU}

\date{\today}

\begin{abstract}

We study the scaling limit of a fully packed loop model in two dimensions,
where the loops are endowed with a bending rigidity. The scaling limit is
described by a {\em three-parameter} family of conformal field theories,  which
we characterize via its Coulomb-gas representation. One choice for two of  the
three parameters reproduces the critical line of the exactly solvable
six-vertex model, while another corresponds to the Flory model of polymer
melting. Exact central charge and critical exponents are calculated for
polymer melting in two dimensions. Contrary to predictions from mean-field
theory we show that polymer melting, as described by the Flory model, is {\em
continuous}. We test our field theoretical results against numerical transfer
matrix calculations.

\end{abstract}


\maketitle

\section{Introduction}

Over the years, polymers physics has greatly benefited from studies of
lattice models. One persistent theme has been the use of lattice models to uncover
universal properties of chain molecules. An example is provided by the scaling exponents which
characterize the statistical properties of polymer conformations, in the limit of very long
chains \cite{degennes}. For polymer chains confined to live in two dimensions, exact values of
exponents were calculated by Nienhuis \cite{nienhuis82} using the self-avoiding walk on the
honeycomb lattice. The predicted value of the swelling exponent, which relates the linear size of the
polymer to the number of monomers, was directly measured in recent fluorescence microscopy
studies of DNA absorbed on a lipid bilayer \cite{DNA}.

Here we turn to the problem of polymer melting, which deals with a  possible phase transition
induced by the competition between chain entropy and bending rigidity. Bending rigidity
determines the persistence length of the polymer. This is the distance over which the relative
orientations of two chain segments are decorrelated due to thermal fluctuations. The long chain
limit mentioned in the previous paragraph is obtained when the polymer
length is much greater than its persistence length.

It is important to point out that the effect of finite  bending rigidity depends
crucially on the steric constraints imposed on the polymer by its interactions with the
solvent. {}For example, in the presence  of a good solvent the polymer is in a ``dilute'' phase.
Typical chain conformations are swollen with empty space between the monomers filled by solvent molecules.
On the lattice, the dilute phase is characterized by a  vanishing fraction of sites occupied
by monomers. In this phase, the bending rigidity simply increases the persistence length of the
polymer, and it does not lead to a  phase transition. This can be verified analytically in two
dimensions, within the framework of Nienhuis' self-avoiding walk model \cite{nienhuis89a,nienhuis89b}.

The picture changes considerably when the polymer is in a ``compact''
phase, with the monomers occupying all the available space. Such a
situation is relevant, for instance, when modelling the conformations
of globular proteins \cite{dill}. Compactness in this case follows from the
interaction between hydrophobic amino-acids and the solvent (water),   which leads to the
expulsion of the solvent from the bulk of the protein. The simplest way to model this
effect is to enforce compactness as a global, steric constraint on the polymer configurations \cite{dill}.
Within this compact phase, one expects a phase transition from a disordered melt to an ordered
crystal as the stiffness of the polymer is increased.

\begin{figure}
\begin{center}
 \leavevmode
 \epsfysize=130pt{\epsffile{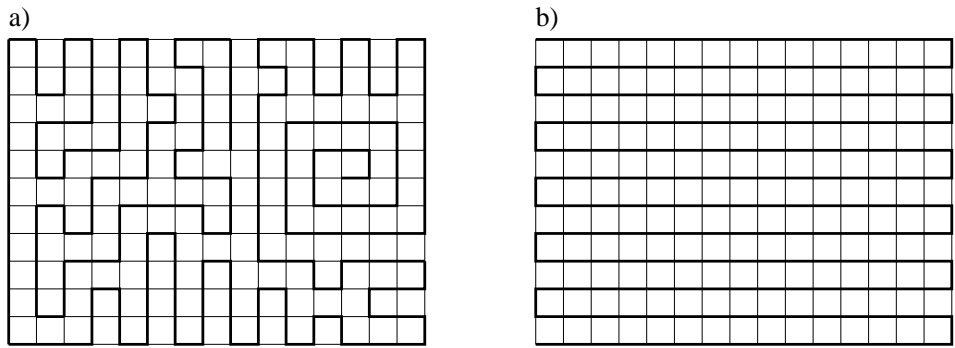}}
 \end{center}
 \protect\caption[3]{\label{fig:chain} Compact polymer configurations on an
 $11 \times 15$ square lattice:  a) Typical configuration in the melt phase,
 and b) zero-temperature  crystalline state, in which the number of bends is
 minimum.}
\end{figure}

To study
this melting transition, in 1956 Flory introduced a lattice model \cite{flory}. Flory's model,
in its simplest formulation, consists of a single chain, described by a self-avoiding walk on the
square lattice, endowed with a bending rigidity. To describe the melted phase the chain is taken to be maximally
compact, filling all the sites of the square lattice; see Fig.~\ref{fig:chain}.
The resistance to bending is modelled by an energy penalty for making $90^{\circ}$
turns.

In the Flory model, at infinite temperature the entropy dominates and the polymer will exhibit a finite
density of bends, as in Fig.~\ref{fig:chain}a. As the temperature is lowered to zero all the
bends are expelled from the bulk and their density goes to zero, as in
Fig.~\ref{fig:chain}b. The nature of the transition from the high temperature melt to the low
temperature crystal has been debated over the years \cite{menon}. Here we show that
the melting transition is {\em continuous} and calculate exact values of scaling
exponents at the transition.

In his original paper, Flory \cite{flory} proposed a mean-field
treatment which predicts a first order transition. According to
\cite{flory}, the density of bends goes to zero at the transition and
the chain entropy vanishes. This prediction of a first order transition with a vanishing entropy
 was challenged by Nagle \cite{nagle}. Namely, he showed that
the exactly solvable six-vertex model maps to a related polymer
model which differs from Flory's by the presence of
polymer loops of all sizes. Applying Flory's mean-field
approximation to this model leads once again to the prediction of
a first order melting transition. However, as Nagle pointed out,
this is at odds with the exact
solution of the six-vertex model \cite{Baxter} which predicts a
continuous, infinite order transition. This observation makes it
questionable that the Flory approach is valid in the original model as well.
In fact, a few years later
Gujrati and Goldstein \cite{gujrati} proved that the polymer entropy
in Flory's model stays finite all the way down to zero temperature
when it finally vanishes. However, the order of the transition still
remained an unresolved question.

Monte Carlo simulations of Baumgartner and Yoon \cite{yoon}, where they allowed
for many chains and a finite density of empty sites, showed a first order
melting transition. Soon thereafter  Saleur \cite{saleur}, using a transfer
matrix approach, presented numerical evidence of a continuous transition,
similar to the one found in the six-vertex model. More recently, Bascle, Garel
and Orland \cite{orland} proposed an improved mean-field treatment of the Flory
model, which does not suffer from the problem of a vanishing entropy at the
transition. It also predicts a first order transition.

Here we show that polymer melting is continuous, as originally argued
by Saleur \cite{saleur}, by making use of a particular model, the
{\em semiflexible loop} (SFL) {\em model}, and its height
representation. Furthermore we calculate the central charge and
exact scaling exponents at the transition. These results are checked
against detailed numerical transfer matrix computations.

The SFL loop model can be thought of as a ``loop generalization'' of the
so-called F-model \cite{nagle}, in which suitably defined loops carry
additional Boltzmann weights. The F-model is a special case of the six-vertex
model \cite{Baxter}, in which all vertices carry equal weights.
This connection will serve as the motivation
for introducing a more general model, the {\em generalized six-vertex
model}, in which the general (zero-field) six-vertex model is endowed with
extra loop weights. We shall finally introduce a similarly generalized
version of the eight-vertex model \cite{Baxter}. Its interest from a
polymer point of view is that it allows for a unified description of
semiflexible lattice polymers in a variety of phases: compact, dense
and dilute. Furthermore it allows us to discuss the effect of vacancies on
the polymer melting transition.

The paper is organized as follows. In the next section we introduce the SFL
model, which, in the limit of zero loop weight, gives the Flory model of
polymer melting, and we discuss its phase diagram. In Sec.~\ref{sec:heights}
we discuss the height representation of the loop model and how it leads to a
conformal field theory in the scaling limit. We make use of the field theory
in Sec.~\ref{sec:ops} to calculate the central charge and scaling exponents,
which we check against numerical transfer matrix computations in
Sec.~\ref{TM_sec}. In Sec.~\ref{sec:pdiagram} we propose a phase diagram for
the generalized six-vertex and eight-vertex models. We end with a
discussion of the scaling of semiflexible compact polymers, and we argue that
the generalized eight-vertex model furnishes a rather complete description of
non-compact semiflexible polymers. An appendix is reserved for a detailed
discussion of the construction of the transfer matrices.

\section{Semiflexible loop model}

Here we define the SFL model, and give a rough sketch of its phase
diagram based on the limits of weak and strong bending rigidity. The
fact that the SFL model reduces to the F-model in the limit of unit
loop fugacity \cite{nagle}, plays an important role in guiding our
intuition about the loop model. It also provides an exactly solvable
line in the phase diagram, against which the field theoretical and
numerical results can be checked.

\subsection{Definition of the model}

The semiflexible fully packed loop model on the square lattice (the
``semiflexible loop model'', or SFL for short) is defined by filling
the square lattice with loops drawn along the lattice edges. Allowed
loop configurations satisfy two constraints:
\begin{itemize}
 \item Self avoidance --- loops are not allowed to cross, and
 \item Full packing --- every site is visited by exactly one loop.
\end{itemize}
On the square lattice
with periodic boundary conditions, edges that are not covered by loops also
form loops, as there are two unoccupied  edges associated with every site of
the lattice. These we refer to as ``ghost loops''.

Given the configurations of the semiflexible loop model, the Boltzmann weights
are defined in the following way. Every real loop is given weight $n_{\rm b}$,
and every ghost loop has weight $n_{\rm g}$. (In all the figures the real and
ghost loops are shown as black and gray respectively, whence the subscripts
${\rm b}$ and ${\rm g}$.) The parameters $n_{\rm b}$ and $n_{\rm g}$ act as
fugacities of the two loop flavors, and as such they control the average
number of loops of each flavor
\footnote{We shall usually consider $n_{\rm b}$ and $n_{\rm g}$ real
and non-negative, though we believe the model to have interesting
properties for negative, or even complex, fugacities.}.  They can be
varied independently as the number of ghost loops is not fixed by the
number of real loops \cite{JK_npb}.  Furthermore, a weight $w_{\rm X}$
is assigned to each vertex of the lattice at which the real and ghost
loops cross. For $w_{\rm X} > 1$ this has the effect of disfavoring vertices
at which the loop makes a 90${}^\circ$ bend, or, in other words, the
loops are semiflexible. The partition function of the semiflexible
loop model is
 \be{partition} Z = \sum_{\cal G} {n_{\rb}^{N_{\rb}}} {n_{\rg}^{N_{\rg}}}
 w_{\rm X}^{V}  \ ,
 \ee
where the sum runs over all allowed loop configurations ${\cal G}$.
$N_{\rm b}$ and $N_{\rm g}$ are the number of real and ghost loops,
respectively, while $V$ is the number of crossing vertices;
these are the two rightmost vertices in Fig.~\ref{fig:6v}. In
the limit $n_{\rb}\to 0$, with $n_{\rg}=1$, we recover the Flory model:
$Z/n_{\rb}$ counts compact polymer loops each weighed by $w_{\rm X}^{V}$.

The semiflexible loop model can be thought of as the generalization of
the \F2 model introduced in Ref.~\cite{JK_npb}. The \F2 model is given
by the partition function, \Eq{partition}, with $w_{\rm X}=1$. It has a
critical phase for $|n_{\rb}|, |n_{\rg}| \le 2$, characterized by a
power law distribution of loop sizes. For other values of the loop
weights the model is non-critical with a distribution of loop sizes
cut off at a finite value (fixed by the correlation length).  Below we
will show that the vertex weight $w_{\rm X}$, for each point in the critical
phase of the \F2 model, produces a line of fixed points which
terminates in a Kosterlitz-Thouless transition.

\subsection{Qualitative phase diagram}

Rough, qualitative features of the phase diagram of the semiflexible
loop model can be deduced from the limits of zero and infinite bending
rigidity. The motivation for developing a precise theory of the phase
diagram, as mentioned in the introduction, stems from the interest in
the $n_{\rm b} \to 0$, $n_{\rm g}=1$ case, which is the Flory model of
polymer melting. We are also motivated by the relation of the SFL
model to the integrable six-vertex model, and its generalizations.

\subsubsection{Flory model}

In the Flory limit of the SFL model, the $w_{\rm X}=1$ point is the
compact polymer problem, which we have studied previously
\cite{JK_npb}. Here one is concerned with enumerating all
self-avoiding walks that visit every site of the lattice. We have
shown that compact polymers on the square lattice are a critical
geometry characterized by non-mean-field scaling exponents which can
be calculated exactly from a field theory.

As $w_{\rm X}$ is increased away from one, we are dealing with a
compact polymer with a bending rigidity. In the limit $w_{\rm X} \to
\infty$ we arrive at a frozen phase in which the density of vertices
at which the polymer bends goes to zero. This is the polymer
crystal. At an intermediate weight $w_{\rm X}=w^{\rm c}_{\rm X}$
($1 < w^{\rm c}_{\rm X} < \infty$) there will be a melting
transition. One of the important unresolved problems is the nature of
this transition. Here we construct an effective field theory of the
Flory model and show that the melting transition is {\em continuous}.

Another interesting issue is the region of $0 \le w_{\rm X} < 1$.
As $w_{\rm X} \to 0$ straight-going vertices are completely suppressed,
and with appropriate boundary conditions the only allowed configurations
are those of a checkerboard pattern of small loops, each loop having its
minimal length of four. If the Flory limit ($n_{\rm b} \to 0$) is taken
before the $w_{\rm X} \to 0$ limit, there has to be a number of straight-going
vertices at the boundary, the dominant configurations being those of a
single wiggly line. In any case, the $w_{\rm X} \to 0$ limit is again
a crystalline phase of zero entropy. We shall however argue below that the
corresponding crystallization transition is located at $w_{\rm X} = 0$ and is thus
rather uninteresting.

\begin{figure}
\begin{center}
 \leavevmode
 \epsfysize=120pt{\epsffile{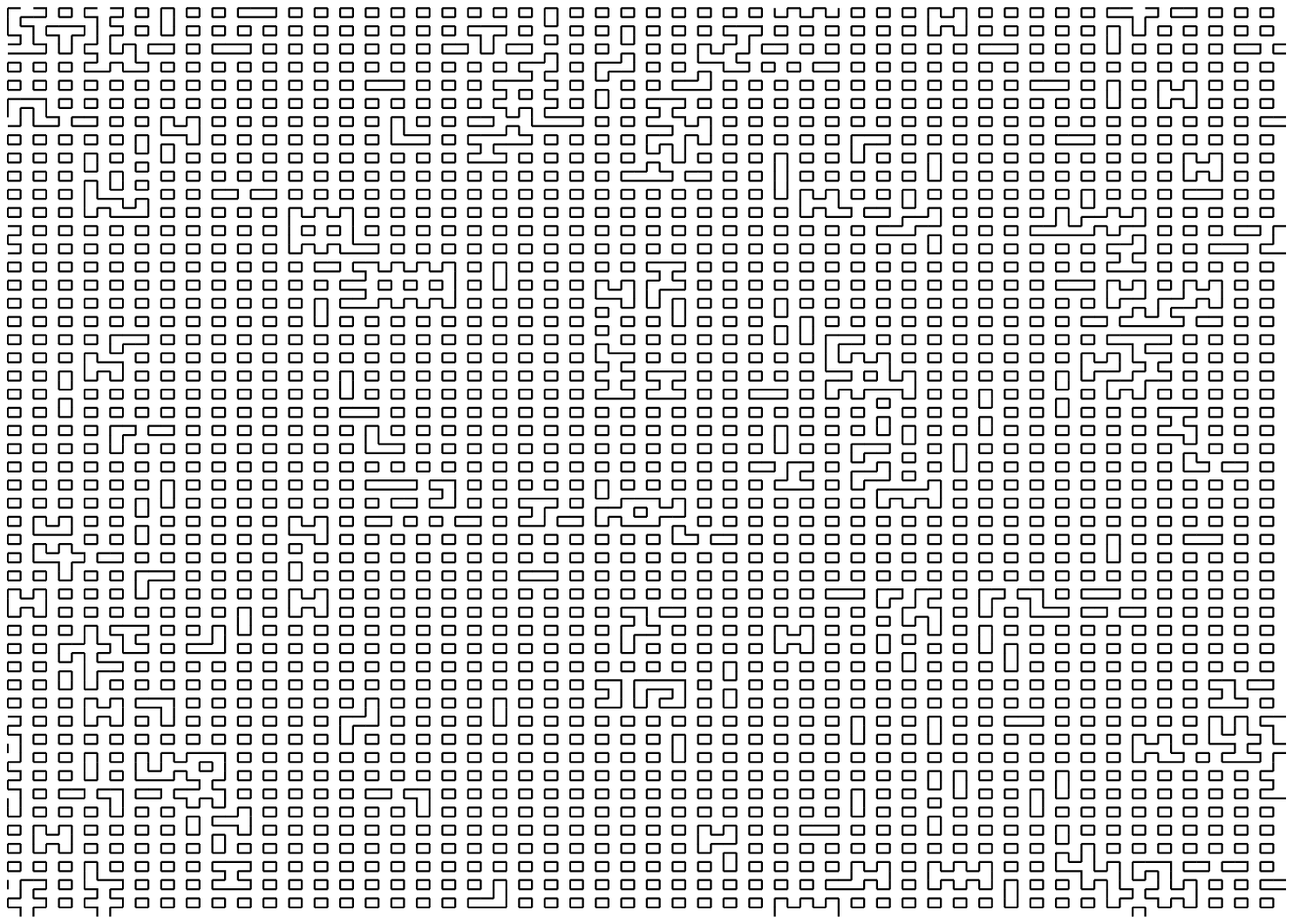}}
 \epsfysize=120pt{\epsffile{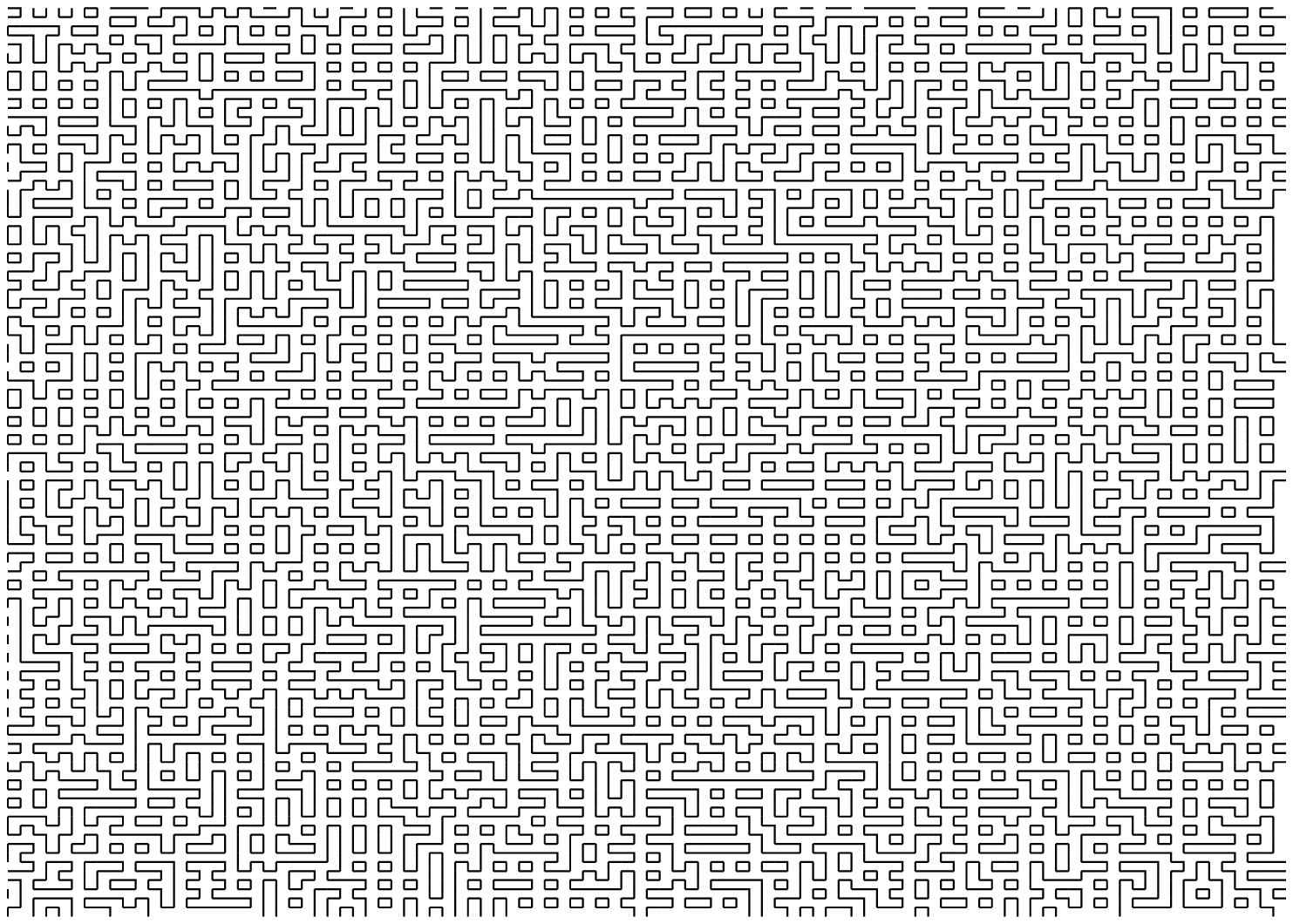}}
 \epsfysize=120pt{\epsffile{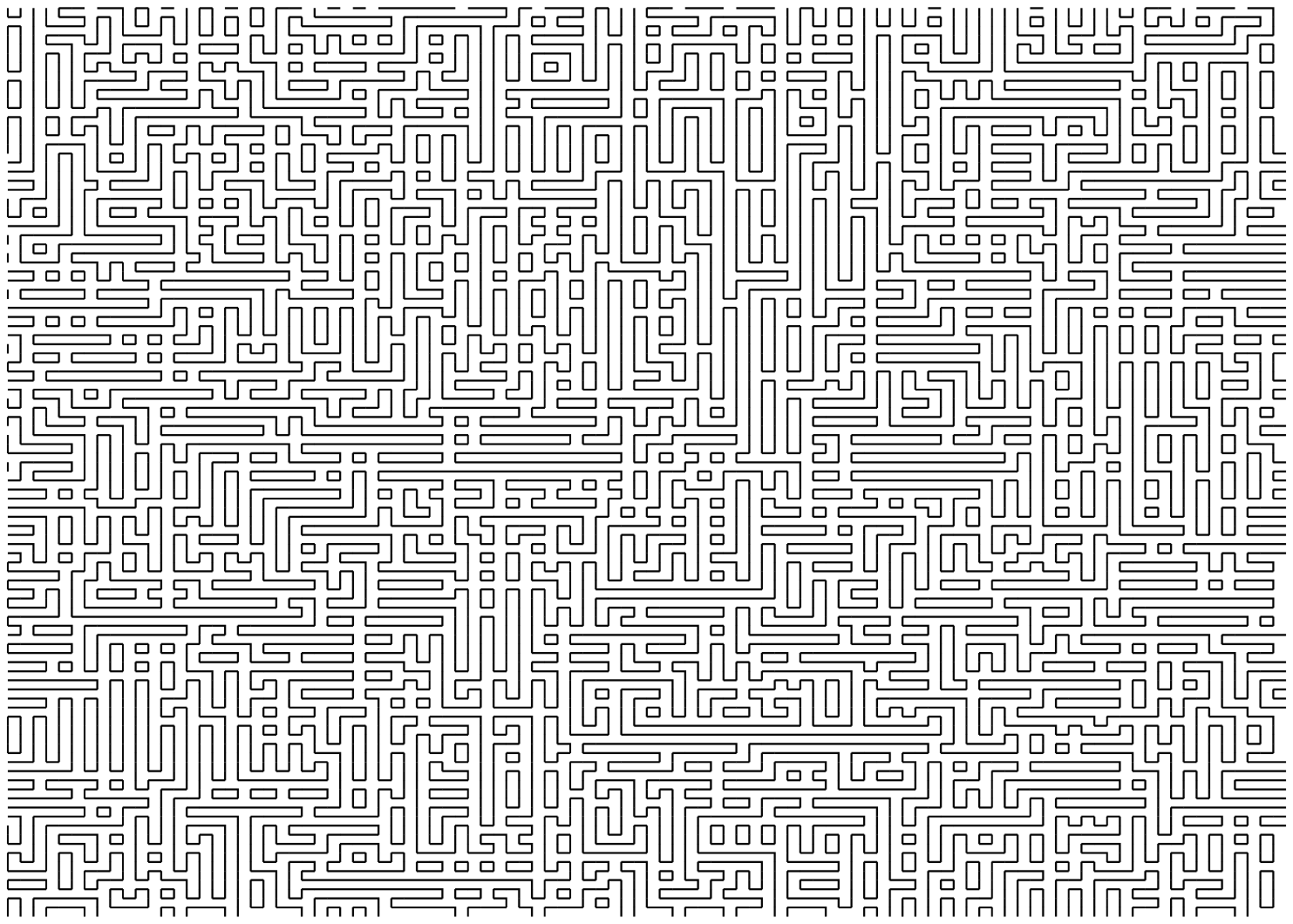}}
 \end{center}
 \protect\caption[3]{\label{fig:config}Typical configurations in the SFL
 model with $n_{\rm b}=n_{\rm g}=2$ and bending rigidity parameter
 $w_{\rm X} = 1/4$ (left panel), $w_{\rm X}=1$ (middle) and
 $w_{\rm X} = 4$ (right). We shall show that the left and middle panels
 correspond to critical melt states, while the right panel is a
 non-critical crystalline state. In the latter, domains of non-zero
 staggered polarization (see Sec.~\ref{6Vm}) are clearly visible.}
\end{figure}

A qualitative idea of the physics underlying the phase diagram of the SFL model
can be obtained by looking at some typical configurations for various values
of $w_{\rm X}$; see Fig.~\ref{fig:config}. The images were obtained
by performing Monte Carlo simulations on a square lattice of size
$100 \times 100$ with toroidal boundary conditions. For technical reasons
\cite{JaneMC} we take $n_{\rm b}=n_{\rm g}=2$ and no loops of
non-contractible topology are allowed. (Further details on the algorithm
used for these simulations can be found in \cite{JaneMC}.)

\subsubsection{Six-vertex model}
\label{6Vm}

\begin{figure}
\begin{center}
 \leavevmode
 \epsfysize=100pt{\epsffile{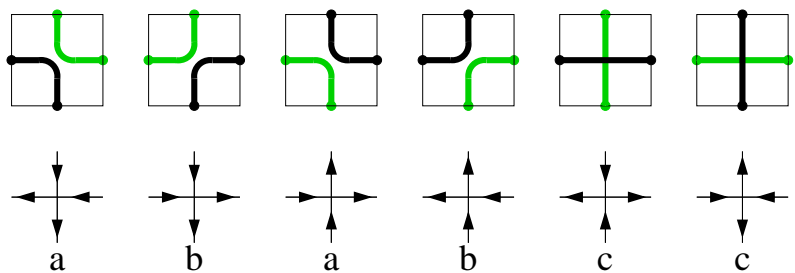}}
 \end{center}
 \protect\caption[3]{\label{fig:6v}Correspondence between the vertices of
 the six-vertex model and the FPL${}^2$ model (here shown for an even vertex;
 at odd vertices the arrows are reversed).}
\end{figure}

Before turning our sights to the semiflexible loop model it is
instructive to review exact results for the (zero-field) six-vertex
(6V) model. The 6V model corresponds to the $n_{\rm b}=n_{\rm g}=1$
line in the phase diagram of the SFL model. The mapping between the
two is simple: at even (odd) vertices the edges covered by the real
loops are identified with arrows pointing out (in), while the edges
covered by the ghost loops correspond to arrows pointing in (out); see
Fig.~\ref{fig:6v}. The appropriate six-vertex weights are: $a=b=1$ and
$c=w_{\rm X}$
\footnote{The vertex weights of the six-vertex model are conventionally
called $a$, $b$ and $c$. We shall henceforth trade the variable $c$ for
$w_{\rm X}$ to avoid confusion with the central charge, which will play
a major role in what follows.}.

In the 6V model there is an order-disorder transition as a function of
the vertex weight $w_{\rm X}$. In the ordered state, which is obtained
for $w_{\rm X} \to \infty$, all the vertices are of the c-variety
(cf.~Fig.~\ref{fig:6v}). The order parameter is the staggered
polarization, which in the loop language can be expressed as the
difference between the number of horizontal and the number of vertical
loop-covered edges, per site \cite{saleur}. The exact solution of the
six-vertex model predicts a continuous (infinite-order) transition
occurring at $w_{\rm X}=2$ \cite{Baxter}. The disordered phase for $0
< w_{\rm X}<2$ is critical with an infinite correlation length and
power-law correlations. Below we will show that there is an analogous
transition in the semiflexible loop model, as $w_{\rm X}$ is varied,
for all values of $|n_{\rm b}|, |n_{\rm g}| \le 2$, including the
Flory case ($n_{\rm b} \to 0, n_{\rm g}=1$). In the Flory model this
was observed previously by Saleur in numerical transfer matrix
computations \cite{saleur}. For the critical phase of the model we
shall construct an effective field theory using the interface
representation of the loop model. This leads to exact (but
non-rigorous) results for the central charge and scaling dimensions,
which we confirm via numerical transfer matrix calculations.

\section{Field theory construction}
\label{sec:heights}

To construct a field theory for the critical phase of the semiflexible
loop model (SFL) we make use of the height representation of the fully
packed loop model on the square lattice (\F2 ). This was already
described in detail in our previous work \cite{JK_npb}, and here it is
briefly reviewed for completeness. The main effect of the vertex
weight $w_{\rm X}$ on the field theory is to renormalize one of its
coupling (elastic) constants. This does not change the central charge,
but leads to continuously varying scaling dimensions for a specific
subset of operators, which we identify. These results are confirmed by
our numerical transfer matrix computations.

\subsection{Height map}
\label{sec:height_map}

The height mapping is defined on the space of oriented loop
configurations $\{ {\cal G}'\}$. We associate $2^{N_{\rm g}+N_{\rm b}}$
oriented loop configuration ${\cal G}'$ with each loop
configuration ${\cal G}$ of the SFL model by independently orienting
every real and every ghost loop clockwise or counterclockwise.

The Boltzmann weight of an oriented loop is $\exp({i \phi})$, where
the phase $\phi=\pm \pi e_{\rm b}$ for clockwise (counterclockwise)
oriented real loops, and $\phi=\pm \pi e_{\rm g}$ for the two
orientations of the ghost loops. To recover $n_{\rm b}$ and $n_{\rm g}$
for the loop weights, after summing over the two possible orientations
we must set
 \ba{fug}
 n_{\rm b} & = & 2 \cos(\pi e_b) \nonumber   \\
 n_{\rm g} & = & 2 \cos(\pi e_g) \ \ .
 \ea
This particular partition of the loop weights between the two
orientations has the advantage of allowing the loop weights to be
distributed among all the vertices that the loop visits, thus
rendering the weights local. This is achieved by assigning the phase
$\pi e_{\rm b}/4$ ($\pi e_{\rm g}/4$) to every vertex at which the
oriented real (ghost) loop makes a right turn, and the opposite phase
for left turns. The fact that for every closed loop on the square
lattice the difference between the number of left and right turns is
$\pm 4$, is what makes these vertex weights work.  The total vertex
weight is then given by the product of phase factors when the loops
bend, while a weight $w_{\rm X}$ is assigned to vertices at which the loops
do not bend.

Turning back to the height map, we define microscopic heights ${\bf
h}({\bf x})$ on the lattice $\{ {\bf x}\}$ dual to the square lattice
on which the loops are defined. Once the height at the origin is
fixed, the heights on all the other vertices of the dual lattice are
uniquely specified by the oriented loop configuration. Namely, the
height difference between nearest neighbor vertices of the dual
lattice is ${\bf A}$, ${\bf B}$, ${\bf C}$ or ${\bf D}$, depending on
the state of the edge that separates them.  The four height-difference
vectors, also referred to as ``colors'', are associated with the four
possible states of any given edge, which can be either covered by a
real or a ghost loop, with one of two possible loop orientations.
Real loops are formed by alternating cycles of ${\bf A}$ and ${\bf B}$ colored
edges, while the ${\bf C}$ and ${\bf D}$ colored edges are ones visited by the
ghost loops. Note that the difference between an ${\bf ABAB}\cdots$ and a
${\bf BABA}\cdots$ cycle encodes the orientation of the corresponding (real)
loop.

The fully packing constraint and the requirement that the height be
unique (i.e., the sum of height differences along any closed lattice path must
be zero), imposes a single algebraic constraint on the four colors:
${\bf A}+{\bf B}+{\bf C}+{\bf D}=0$. It follows that only three of the
four vectors are linearly independent. A convenient choice that
respects the symmetries between the four colors is to let the
corresponding vectors point from the center to the vertices of a
regular tetrahedron:
\begin{eqnarray}
  {\bf A} = (-1,+1,+1) \ , \ \ \ \ {\bf B} = (+1,+1,-1) \ , \nonumber \\
  {\bf C} = (-1,-1,-1) \ , \ \ \ \ {\bf D} = (+1,-1,+1) \ .
  \label{colours1}
\end{eqnarray}

The effective field theory for the SFL model describes the
fluctuations of the coarse-grained heights which retains only the
long-wavelength (much larger than the lattice spacing) Fourier modes
of the microscopic heights.

\subsection{Effective field theory: $w_{\rm X}=1$}

For $w_{\rm X}=1$ we have the familiar case of the fully packed loop
model on the square lattice. Its effective field theory was discusses
in a previous publication \cite{JK_npb} and here it is reviewed for
completeness.

The partition function of the loop model in the height representation can be
written as a path integral over the coarse grained heights with the
(dimensionless) action:
\be{action_tot}
 S = S_{\rm E} + S_{\rm B} + S_{\rm L} \ .
\ee
This action only takes into account the long-wavelength fluctuations of the
microscopic height. The three terms in the action are of different origin.

The elastic term,
\be{action_el}
S_{\rm E} =
 \frac{1}{2} \int \! {\rm d}^2{\bf x} \ \left \lbrace
      K_{11}[(\bp h^1)^2 + (\bp h^3)^2]\right. +
      2 K_{13}(\bp h^1 \cdot \bp h^3) +  \left. K_{22} (\bp h^2)^2
      \right \rbrace \  ,
\ee
accounts for the height fluctuations due to the entropy of fully
packing the square lattice with oriented loops. Equivalently, this is
the entropy of edge coloring the square lattice with four different
colors. The elastic term favors oriented loop configurations that
minimize the variance of the microscopic height; these are the
macroscopically flat states.  In terms of the color degrees of freedom
the flat states have the property that the four edges of each
elementary plaquette are colored by two colors only.

The particular form of the matrix of elastic constants, ${\bf K}$, is
fixed by the lattice symmetries and symmetries associated with
permuting the colors ${\bf A}$, ${\bf B}$, ${\bf C}$, and ${\bf
D}$. The elastic constants $K_{ij}$ are functions of the loop
fugacity. For the $w_{\rm X}=1$ case they were calculated in
Ref.~\cite{JK_npb} using the loop ansatz \cite{JK_prl}, which allows
one to identify the marginal screening charges \cite{dots}.  For the
\F2 model there are four screening charges:
\ba{marg_ch}
{\bf e}^{(1)} & = & (-\pi, 0, +\pi) \nonumber \\
{\bf e}^{(2)} & = & (-\pi, 0, -\pi) \nonumber \\
{\bf e}^{(3)} & = & (-\pi, +\pi, 0)  \nonumber \\
{\bf e}^{(4)} & = & (-\pi, -\pi, 0) \ .
\ea
These electric charges are associated with the most relevant vertex
operators appearing in the Fourier expansion of the operator conjugate
to the loop weight (see \Eq{F_sum2}, below).

Demanding that all four charges have scaling dimension equal to two gives (using Eq.~\ref{em_dim})
\ba{el_cst}
K_{11} & = & \frac{\pi}{8} (2-e_g-e_b)   \nonumber \\
K_{13} & = & \frac{\pi}{8} (e_b-e_g) \\
K_{22} & = & \frac{\pi}{2} \frac{(1-e_b)(1-e_g)}{2-e_b-e_g}  \nonumber
\ea
for the elastic constants of the FPL${}^2$ model; $e_{\rm b}$ and
$e_{\rm g}$ satisfy \Eq{fug} and take their values on the interval
$[0, 1/2]$.  Below we will argue that the effect of $w_{\rm X}\neq 1$ is to
change the value of the elastic constant $K_{22}$ while leaving the
other two unchanged.

The boundary term in the action,
\be{action_bc}
 S_{\rm B} = \frac{{\rm i}}{4 \pi} \int \! {\rm d}^2{\bf x} \ ({\bf e}_0
             \cdot {\bf h}({\bf x})) \rho({\bf x}) \ ,
\ee
enforces the correct weight of topologically non-trivial loops. If the
oriented loop model is defined with periodic boundary conditions along
one direction (i.e., on a cylinder) these would be the loops that
completely wind around the cylinder
\footnote{Note in particular that the field theory construction only
applies to geometries that are isomorphic to the sphere with a point at
infinity under a conformal transformation. Generalizations to higher
genera have so far only been treated in mean-field theory
\cite{cyrano,higuchi}.}.
On a cylinder the scalar curvature, $\rho$, is non-zero only at the
two boundaries at infinity. $S_{\rm B}$ has the effect of placing
background electric charges $\pm{\bf e}_0$ at the two boundaries,
where the identification
\be{bc}
 {\bf e}_0 = - \frac{\pi}{2} (e_{\rm g} + e_{\rm b}, 0, e_{\rm g} - e_{\rm b}) \
\ee
comes about by demanding that the oriented winding loops be assigned correct
phase factors, $\exp(\pm {\rm i} \pi e_{\rm b})$ or $\exp(\pm {\rm i} \pi e_{\rm g}$) \cite{JK_npb}.

The third term, called the Liouville term,
 \be{action_lp} S_{\rm L} = \int \! {\rm d}^2{\bf x} \
   w[{\bf h}({\bf x})] \ ,
 \ee
owes its existence to the complex weights associated with oriented
loops in the bulk. The local redistribution of the loop weights made
in Sec.~\ref{sec:height_map} leads to complex vertex weights,
which in turn depend only on the colors of the four edges around the
vertex. If we write the vertex weight as $\exp(-w)$ then
\ba{weight_op}
w({\bf B}, {\bf C}, {\bf A}, {\bf D}) & = &  0, \nonumber \\
w({\bf B}, {\bf D}, {\bf A}, {\bf C}) & = &  0, \nonumber \\
w({\bf A}, {\bf B}, {\bf C}, {\bf D}) & = & \mp {\rm i} \frac{\pi}{4}
                                           (e_{\rm g}+e_{\rm b}), \nonumber \\
w({\bf B}, {\bf A}, {\bf C}, {\bf D}) & = & \mp {\rm i} \frac{\pi}{4}
                                           (e_{\rm g}-e_{\rm b}), \nonumber \\
w({\bf A}, {\bf B}, {\bf D}, {\bf C}) & = & \mp {\rm i} \frac{\pi}{4}
                                           (e_{\rm b}-e_{\rm g}), \nonumber \\
w({\bf B}, {\bf A}, {\bf D}, {\bf C}) & = & \mp {\rm i} \frac{\pi}{4}
                                           (-e_{\rm b}-e_{\rm g}) \ ;
\ea
the top (bottom) sign is for even (odd) vertices, and the colors are
listed in order, starting from the left-most edge and proceeding
clockwise around the vertex. The weight operator $w$ is invariant
under cyclic permutations of the colors and it is a periodic function
of the heights around a vertex. In the scaling limit the vertex
weights give rise to the operator $w[{\bf h}({\bf x})]$ in
\Eq{action_lp} which can be written as a Fourier series \be{F_sum2}
w[{\bf h}({\bf x})] = \sum_{{\bf e}\in{\cal R}^*_w} \tilde{w}_{\bf e}
\exp({\rm i} {\bf e} \cdot {\bf h}({\bf x})) \ .  \ee The electric
charges ${\bf e}$ appearing in the Fourier expansion are dictated by
the lattice of periodicities ${{\cal R}_w}$ of the operator $w[{\bf
h}]$; ${{\cal R}^*_w}$ is the reciprocal lattice.  ${{\cal R}_w}$ is
determined by inspection of the values the loop weight operator takes
on the flat states: vectors in ${{\cal R}_w}$ connect flat states on
which the loop weight operator takes identical values.  The most
relevant charges in ${{\cal R}^*_w}$ are the four given in
\Eq{marg_ch}. We identify them with the screening charges \cite{dots}
of the Coulomb gas.  This is the content of the loop ansatz introduced
in Ref.~\cite{JK_prl}.

\subsection{Effective field theory: $w_{\rm X}\neq1$}

For the SFL model, when $w_{\rm X}\neq1$, the Liouville term in
\Eq{action_tot} is modified, while the elastic and the boundary terms
are unchanged. The number of marginal screening charges appearing in
\Eq{F_sum2} is reduced from four to two, and the loop ansatz fixes the
values of $K_{11}$ and $K_{13}$ only. They do not depend on the value
of $w_{\rm X}$ and are given by the $w_{\rm X}=1$ formulae,
\Eq{el_cst}. $K_{22}$, on the other hand, is a non-universal function
of $w_{\rm X}$. Below we present arguments for this scenario, which is
supported by exact results available in the 6V case (i.e., for $n_{\rm
b}=n_{\rm g}=1$), and by our numerical transfer matrix calculations
described in Sec.~\ref{TM_sec} below.

The new vertex weight $w_{\rm X}$ changes the value of $w$ in
\Eq{weight_op} from $0$ to $-\ln w_{\rm X}$ for the vertex states
$({\bf B},{\bf C},{\bf A},{\bf D})$, $({\bf B},{\bf D},{\bf A},{\bf
C})$, and six other related to these two by cyclic permutations of the
colors. The weights of the other 16 vertex states are unchanged. We
consider the consequences of this change on the effective field
theory.

In the height representation of the SFL model, the change in vertex weight
corresponds to adding
\be{Sx}
  S_{\rm X} = \int \! {\rm d}^2{\bf x} \ X[{\bf h}({\bf x})] \ ,
\ee
to the action. The $X$ operator takes the value $\ln w_{\rm X}$ on the flat
states made up of $({\bf B}, {\bf C}, {\bf A}, {\bf D})$ or $({\bf B},
{\bf D}, {\bf A}, {\bf C})$ type vertices, and vanishes on all the
others.  By inspection of the graph of flat states we find that the
lattice of periodicities for the operator $X$, ${{\cal R}_X}$, is the
span of $(1,0,-1)$, $(1,0,1)$ and $(0,1,0)$; these are the height
difference vectors between the flat states in the support of $X$.
This observation implies that $X[{\bf h}]$ can be expanded in a
Fourier series over electric charges that live in the dual lattice
${{\cal R}^*_X}$ which is the span of $(\pi,0,\pi)$, $(\pi,0,-\pi)$
and $(0,2\pi,0)$.

If we consider the effect of $S_{\rm X}$ as a perturbation on the
action of the FPL${}^2$ model the electric charges $(0,\pm 2\pi,0)$
play a special role.
Namely, the operator product expansion of $\exp{(i(0,2\pi,0)\cdot{\bf
h})}$ and $\exp{(-i(0,2\pi,0)\cdot{\bf h})}$ contains the $(\bp
h^2)^2$ operator, and therefore leads to the renormalization of
$K_{22}$ \cite{cardy_book}. This follows from the fact that the
background charge, \Eq{bc}, has a vanishing second component. On the
other hand, for charges ${\bf e}$ with non-zero first or third
component, the effect of the background charge is  that the operator
product expansion of $\exp{(i{\bf e}\cdot{\bf h})}$ and $\exp{(-i{\bf
e}\cdot{\bf h})}$ does not contain $\bp h^i \cdot \bp h^j$ operators and
therefore does not lead to the renormalization of the elastic
constants $K_{ij}$.

The Coulomb gas representation of the height model provides a clear
physical picture of the effect of $S_{\rm X}$ on the critical action
of the \F2 model.  For $w_{\rm X}=1$ the dimension of the $(0,\pm2\pi,0)$
charges follows from \Eq{em_dim},
\be{dim_X}
 x_X = \frac{(2\pi)^2}{4 \pi K_{22}} =
 2 \left ( \frac{1}{1-e_b} + \frac{1}{1-e_g} \right ) .
\ee
It is greater than 2 in the whole critical region of the FPL${}^2$
model. These charges are therefore irrelevant in the renormalization
group sense. In the Coulomb gas picture the $(0,\pm2\pi,0)$ charges
appear as bound pairs of neutral dipoles. Increasing $w_{\rm X}$ will have
the effect of increasing the bare fugacity of these dipoles which will
in turn increase the value of the coupling $K_{22}$ appearing in the
effective field theory. Formally, this can be seen in perturbation
theory making use of the operator product expansion
\cite{cardy_book}. Physically, the renormalization of $K_{22}$ can be
understood as the screening effect of dipoles. The dipoles lower the
Coulomb energy between two electric test-charges having a non-zero
second component, corresponding to an increase in the value of
$K_{22}$ which plays the role of a dielectric constant.

At a critical value $w^{\rm c}_{\rm X}$ there will be a Kosterlitz-Thouless
type transition of the SFL model into a flat state with a vanishing
density of vertices at which the polymer bends.  At the transition the
$(0,\pm2\pi,0)$ charges are marginal, i.e., their scaling dimension
is equal to 2. Using \Eq{dim_X} this observation gives rise to the
prediction for the critical value of $K_{22}$:
\be{stiff22_c}
 K_{22}(w^{\rm c}_{\rm X}) = \frac{\pi}{2} \ .
\ee
For values of $w_{\rm X}$ smaller than $w^{\rm c}_{\rm X}$, $K_{22}$
will be a {\em non-universal} function of $w_{\rm X}$. In the $n_{\rm
b}=n_{\rm g}=1$ case, the formula $K_{22} = \arcsin(w_{\rm X}/2)$
follows from the exact solution of the 6V model \cite{Baxter}. The
critical value of the vertex weight is $w_{\rm X}^{\rm c}({\rm 6V}) =
2$ and $K_{22}(2) = {\pi}/{2}$ is in agreement with
\Eq{stiff22_c}. For other values of $n_{\rm b}$ and $n_{\rm g}$ our numerical
transfer matrix calculations are in good agreement with
\Eq{stiff22_c}.

The introduction of the vertex weight $w_{\rm X}$ also has an effect on the
screening charges, \Eq{marg_ch}, that appear in the Liouville part of
the action. First consider the $n_{\rm b}=n_{\rm g}$ case of the SFL
model. Due to the presence of the $w_{\rm X}$ term cyclic permutations of
the four colors around a vertex are no longer a symmetry of the vertex
weight. Therefore, unlike the $w_{\rm X}=1$ case \cite{JK_npb}, there are
now two independent elastic constants, $K_{22}$ and $K_{11}$ appearing
in $S_{\rm E}$. $K_{13}=0$ follows from the remaining $Z_2$ symmetry
of the vertex weights which are invariant under {\em two} cyclic
permutations, such as $({\bf A}, {\bf B}, {\bf C}, {\bf D}) \to ({\bf
C}, {\bf D}, {\bf A}, {\bf B})$.  The deduced structure of the
elasticity matrix implies that the four electric charges in
\Eq{marg_ch} are no longer degenerate in dimension for arbitrary
$w_{\rm X}$.  Since the dimensions of ${\bf e}^{(1)}$ and ${\bf e}^{(2)}$
are independent of $K_{22}$ they are identified as the two screening
charges tied to the non-renormalizability of the loop weights
\cite{JK_npb}.  As in the FPL${}^2$ model we then assume that these
two charges remain marginal when $n_b\neq n_g$. Using the dimension
formula, \Eq{em_dim}, this then fixes the values of the two elastic
constants, $K_{11}$ and $K_{13}$, to the values quoted in \Eq{el_cst}.

Finally, it is interesting to look at some extreme limits of $K_{22}$
in view of the effective field theory. Consider first the limit
$K_{22} \to \infty$ in which height fluctuations in the second height
component are completely suppressed. (As we are outside the critical
phase, we are here referring to the bare value of the coupling.)
Clearly, height fluctuations
must always be present in the microscopic four-coloring model, but it
is nevertheless instructive to look for the states that minimize the
fluctuations of $h^2$. From the choice of the color vectors,
Eq.~(\ref{colours1}), it is not difficult to see that on the four
sites of $\{ {\bf x} \}$ surrounding a given vertex, $h^2$ fluctuates
by two units for the first four vertices of Fig.~\ref{fig:6v} and by
one unit for the last two vertices. All vertices must therefore be of
the c-type, corresponding to the limit $w_{\rm X} \to \infty$.  Thus,
$K_{22} \to \infty$ as $w_{\rm X} \to \infty$.

Conversely, as $K_{22} \to 0$ the fluctuations in $h^2$ become unbounded
and the effective field theory loses its consistency (since it was based
on the assumption that the interfacial entropy is due to bounded fluctuations
around the macroscopically flat states). However, the argument given above
indicates that a small value of $K_{22}$ should correspond to a small
number of straight-going vertices in the loop model. Thus, we would
conjecture that $K_{22} \to 0$ as $w_{\rm X} \to 0$. This expectation is
confirmed by the exact result for the 6V case \cite{Baxter} and also by
extrapolation of our numerical results for $K_{22}(w_{\rm X})$ in the
Flory case.

Apart from these limiting values, we would of course expect $K_{22}$ to be
a monotonically increasing function of $w_{\rm X}$ throughout the critical
phase.

In the next section we compute the central charge and the scaling
dimensions of various operators in the semiflexible loop model from
its effective field theory. We identify quantities that depend on the
non-universal elastic constant $K_{22}$; these are then predicted to vary
continuously with $w_{\rm X}$.

\section{Operators and scaling dimensions}
\label{sec:ops}

The effective field theory of the semiflexible loop model describes a
Coulomb gas of electric and magnetic charges in the presence of
background and screening charges.  The magnetic charges ${\bf m}$ are
vectors in ${\cal R}$ which is the lattice of periodicities of the
graph of flat states, while the electric charges ${\bf e}$ take their
values in the reciprocal lattice ${\cal R}^*$ \cite{JK_npb}.  With the
normalization adopted for the vectors ${\bf A}$ through ${\bf D}$,
\Eq{colours1}, ${\cal R}$ is a face-centered cubic lattice whose
conventional cubic cell has sides of length 4, while is ${\cal R}^*$
is a body-centered cubic lattice whose conventional cubic cell has
sides of length $\pi$.

The scaling dimension of an operator which has total electromagnetic
charge $({\bf e}, {\bf m})$ is the sum of its electric and magnetic
dimensions, and it is a function of the elastic constants and the
background charge \cite{dots}:
\be{em_dim}
x({\bf e}, {\bf m}) = \frac{1}{4\pi} \left [ ({\bf e} {\bf K}^{-1}) \cdot
     ({\bf e} - 2 {\bf e}_0) + ({\bf m} {\bf K}) \cdot {\bf m} \right ] \ .
\ee
${\bf K}$ is the $3\times 3$ matrix of elastic constants and ${\bf
K}^{-1}$ is its inverse.

{}From \Eq{em_dim} and the form of ${\bf K}$ (\Eq{action_el}) and
${\bf e}_0$ (\Eq{bc}) it immediately follows that operators whose electric and
magnetic charges both have a vanishing second component will have a
$K_{22}$-independent scaling dimension. The scaling dimension in this
case is independent of $w_{\rm X}$ and equal to its known value at
$w_{\rm X}=1$ \cite{JK_npb}. Operators with ${\bf e}$ and ${\bf m}$  charges whose
second components are not both zero will, on the other hand, have a scaling
dimension that varies continuously with $w_{\rm X}$. These predictions are
confirmed by our numerical results.

\subsection{Central charge}

The central charge of the SFL model follows from its critical
action. The three height components (bosonic free fields) each contribute
one unit to the central charge while the contribution from the background
charge is $12 x({\bf e}_0, 0)$. Using \Eq{em_dim} for $x({\bf e}_0, 0)$, \Eq{bc} for the background
charge ${\bf e}_0$, and the calculated values of the elastic constants
$K_{11}$ and $K_{13}$, \Eq{el_cst}, we find
\be{c_charge1}
 c = 3 - 6\left( \frac{e_{\rm b}^2}{1 - e_{\rm b}}
      + \frac{e_{\rm g}^2}{1 - e_{\rm g}} \right),
\ee
{\em independent} of the unknown value of $K_{22}$.

For the 6V model, which corresponds to the $e_{\rm b}=e_{\rm g}=1/3$
line in the SFL model, the above formula gives $c=1$ for the central
charge along the critical line. This result also follows directly
from the exact solution of the 6V model.

For the Flory model of polymer melting, which is the $e_{\rm b}=1/2$,
$e_{\rm g}=1/3$ case, the predicted central charge is $c=-1$. This value is
confirmed by our numerical transfer matrix calculations (see Sec.~\ref{TM_sec}).

\subsection{Thermal operator}

The SFL model can be thought of as the zero-temperature limit of a
more general model where we allow for thermal excitations that violate
the fully packing constraint. Violations of the constraint lead to
vertices with the four adjacent edges colored $({\bf C},{\bf D},{\bf
C},{\bf D})$. In the height representation such a vertex is identified
with a topological defect (screw dislocation) whose charge, i.e., the sum of height
differences around the vertex, is
\be{mT_strength}
 {\bf m}_T = 2 ({\bf C} + {\bf D}) = (0,-4,0) .
\ee
Other vertices which have no ${\bf A}$ or ${\bf B}$ colored edges are
possible, but they have a larger magnetic charge and are hence less
relevant.

In the Coulomb gas picture a topological defect corresponds to
a magnetic charge. Therefore, the thermal dimension can be calculated
using \Eq{em_dim}, and we find
\be{xT}
  x_T =  x(0,{\bf m}_T) = \frac{4}{\pi} K_{22} \ .
\ee
We make use of this equation below as it allows us to determine the
unknown elastic constant $K_{22}$ from a measurement of the thermal
scaling dimension. Once this elastic constant is known, scaling
dimensions of all electromagnetic operators can be calculated from
\Eq{em_dim}.

\subsection{String operators}
\label{string_sec}

A particularly important set of operators in any loop model are the
string operators. Their two-point function is defined as the
probability of having the small neighborhoods around two fixed points
on the lattice, which are separated by a large distance, connected by
$s_{\rm b}$ real loop segments and $s_{\rm g}$ ghost loop segments. For
simplicity, we shall require $s_{\rm b}$ and $s_{\rm g}$ to be either
both even or both odd; $s_{\rm b}+s_{\rm g}$ odd requires $L$ to
be odd which produces a twist in the height, as discussed in
Ref.~\cite{JK_npb}. In the height representation these string
configurations are mapped to two topological defects, one serving as
the source and the other as the sink of oriented loop segments. When
the oriented loop segments wind around the defect points they are assigned
spurious phase factors by the vertex weights; these phase factors can
however be compensated by introducing appropriate
electric charges at the positions of the defects \cite{nien_rev}.

In the case $s_{\rm b}=2 k_{\rb}$ and $s_{\rm g}=2 k_{\rg}$, i.e.,
when the number of real and ghost strings are both even, the electric
and magnetic charge of the corresponding string operator are
\cite{JK_npb}
\ba{em_even}
{\bf e}_{2k_{\rb},2k_{\rg}} & = & -\frac{\pi}{2} (e_{\rb}, 0, -e_{\rb})(1-\delta_{k_{\rb},0})
-\frac{\pi}{2}(e_{\rg}, 0, e_{\rg})(1-\delta_{k_{\rg},0}) \nonumber \\
{\bf m}_{2k_{\rb}, 2k_{\rg}} & = & -2(k_{\rb}+k_{\rg},0,k_{\rg}-k_{\rb}) \ .
\ea

Since the charges have vanishing second component their dimension is
independent of $K_{22}$ and constant along the whole critical line
$w_{\rm X} \le w^{\rm c}_{\rm X}$. The value of the string dimension
follows from \Eq{em_dim},
\be{x_even}
 x_{2k_{\rb},2k_{\rg}} = \frac{1}{2} \left [ (1-e_{\rm b}) k_{\rb}^2 +
            (1-e_{\rm g}) k_{\rg}^2  -
            \frac{e_{\rm b}^2}{1-e_{\rm b}} (1-\delta_{k_{\rb},0})
         -  \frac{e_{\rm g}^2}{1-e_{\rm g}} (1-\delta_{k_{\rg},0}) \right ]
\ee
and is identical to that of the \F2 model \cite{JK_npb}. Our numerical
simulations confirm that even string dimensions are constant along the
critical line.

In the odd string case, when $s_{\rm b}=2 k_{\rb}-1$ and $s_{\rm g}=2
k_{\rg}-1$, the electric and magnetic charge are \cite{JK_npb}
\ba{em_odd}
{\bf e}_{2k_{\rb}-1,2k_{\rg}-1} & = & - \frac{\pi}{2} (e_{\rm g} + e_{\rm b}, 0, e_{\rm g} - e_{\rm b}) \nonumber \\
{\bf m}_{2k_{\rb}-1,2k_{\rg}-1} & = &  -2 (k_{\rb}+k_{\rg}-1, 1, k_{\rg}-k_{\rb}) \ .
\ea
Notably the magnetic charge has a non-vanishing second
component. Using \Eq{em_dim} we calculate
\be{x_odd}
x_{2k_{\rb}-1, 2k_{\rg}-1} =  \frac{K_{22}}{\pi} +
\frac{1}{8} \left[ (1-e_{\rm b}) (2k_{\rb}-1)^2 + (1-e_{\rm g})(2k_{\rg}-1)^2 \right] - \frac{1}{2}
\left[ \frac{e_{\rm b}^2}{1-e_{\rm b}} +
                \frac{e_{\rm g}^2}{1-e_{\rm g}} \right]
\ee
for the odd string dimension. It depends on the value of $K_{22}$ and
will therefore vary continuously with $w_{\rm X}$. At the melting transition
the exponents are exactly known from \Eq{stiff22_c}. This is
confirmed by our numerical transfer matrix results, which we describe
next.

\section{Transfer matrix results}
\label{TM_sec}

To check the correctness of our field theoretical predictions, we have
numerically diagonalized the transfer matrix of the semi-flexible loop
model (and of its various generalizations, to be discussed below)
defined on semi-infinite cylinders of even widths $L$ ranging from 4 to
14.

The existence of a transfer matrix may not be a priori obvious,
since the Boltzmann weights depend on the number of loops, which is a
non-local quantity. We have however already shown in an earlier
publication \cite{JK_npb} how this is resolved by working in a basis
of states that contains non-local information about how loop segments
are interconnected at a given stage of the computations. In that paper,
it was also shown that the full transfer matrix contains various
sectors, the leading eigenvalues of which provide finite-size
estimations of the free energy and of the various critical exponents,
using the standard CFT relations \cite{central,gaps}
\ba{FSS}
 f_0(L) &=& f_0(\infty) - \frac{\pi c}{6 L^2} + \cdots \\
 f_k(L) - f_0(L) &=& \frac{2 \pi x_k}{L^2} + \cdots \label{FSS_2}
\ea
Here, the label $k$ refers either to a higher eigenvalue in the
sector to which $f_0$ belongs, or to the leading eigenvalue in another
sector characterized by some topological defect of charge $k$.

To access critical exponents, we shall mainly be concerned with
topological defects that consist in enforcing that a certain number
of strings of either flavor propagate along the length direction of
the cylinder. These give rise to a two-parameter family of critical
exponents $x_{s_{\rb},s_{\rg}}$ corresponding to $s_{\rb}$ real strings
and $s_{\rg}$ ghost strings. The corresponding topological charges,
\Eq{em_even} and \Eq{em_odd},
take the form of three-dimensional electromagnetic vector charges.
In the transfer matrix calculations, each of these topological sectors is associated
with a different state space. The difficulty of precisely
characterizing these spaces limited our previous approach \cite{JK_npb} to
at most two strings. In Appendix~\ref{appA} we present an algorithm
that explicitly constructs the required state spaces for any
$(s_{\rb},s_{\rg})$, based on an iterative procedure and hashing techniques.

By inspection of the eigenstates produced by our previous algorithm
\cite{JK_npb}, it turns out that many of the basis states carry zero
weight. One would then expect that identical results can be obtained
more efficiently by working in a basis in which such states have been
eliminated from the outset. We defer the technical details of how this
can be done to Appendix~\ref{appA}. It is also shown how the
block-diagonalization scheme can be carried even further, by
exploiting various conservation laws that are most easily understood
from the analogy between the SFL model and the six-vertex model. One
important consequence is that the constrained free energy $f_T(L)$
that is linked to the thermal scaling dimension can now be obtained as
a leading eigenvalue, rather than as the second eigenvalue in the
stringless sector. This considerably improves the efficiency of the
computations.

Finally, the matrix elements need some modification in order to take
into account the bending rigidity parameter $w_{\rm X}$. This is readily
done, without any modification of the basis states, since $w_{\rm X}$ is a
purely local quantity.

Before turning to our numerical results, we should mention that we
have submitted our transfer matrices to several tests, in order to
verify their correctness:
\begin{itemize}
 \item For $w_{\rm X}=1$, all numerically determined string dimensions $x_{s_{\rb},s_{\rg}}$ with
       $s_{\rb}+s_{\rg}=2$ or $4$ agree to at least three significant
       digits with their exact values in the cases $(n_{\rb},n_{\rg})=(1,1)$
       \cite{Baxter} and $(n_{\rb},n_{\rg})=(0,1)$ \cite{JK_npb}.
 \item All eigenvalues found for the FPL${}^2$ model agree with
       those obtained from our previous algorithm \cite{JK_npb}.
 \item For the six-vertex model \cite{Baxter}, we have compared the
       extrapolated bulk free energy with Baxter's exact expression.
 \item Again for the six-vertex model, we find excellent agreement with the
       exact formulae $x_{1,1} = \frac{K_{22}}{\pi}$ and
       $x_T = \frac{4K_{22}}{\pi}$, where $K_{22}=\arcsin(w_{\rm X}/2)$ is the
       elastic constant.
 \item We have also found agreement with the first few terms in
       diagrammatic expansions around various limits of infinite fugacities.
\end{itemize}

\subsection{Central charge}

A crucial prediction of our field theory is that, for given values of
the loop fugacities $(n_{\rb},n_{\rg})$, the central charge of the SFL model
should be independent of $w_{\rm X}$, as long as the latter is constrained
to the critical regime, $0 < w_{\rm X} \le w_{\rm X}^{\rm c}$.

\begin{figure}
\begin{center}
 \leavevmode
 \epsfysize=130pt{\epsffile{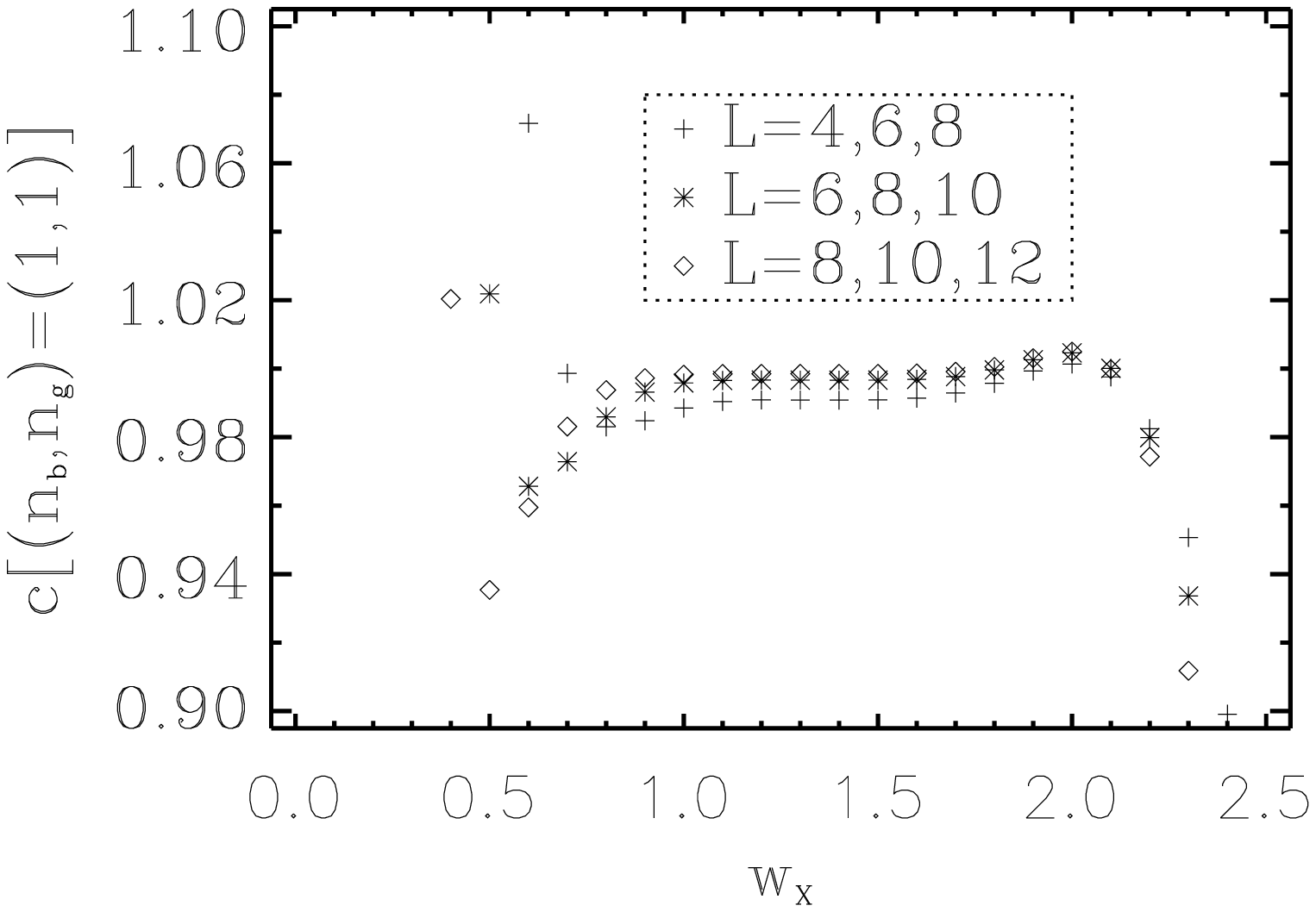}}
 \epsfysize=130pt{\epsffile{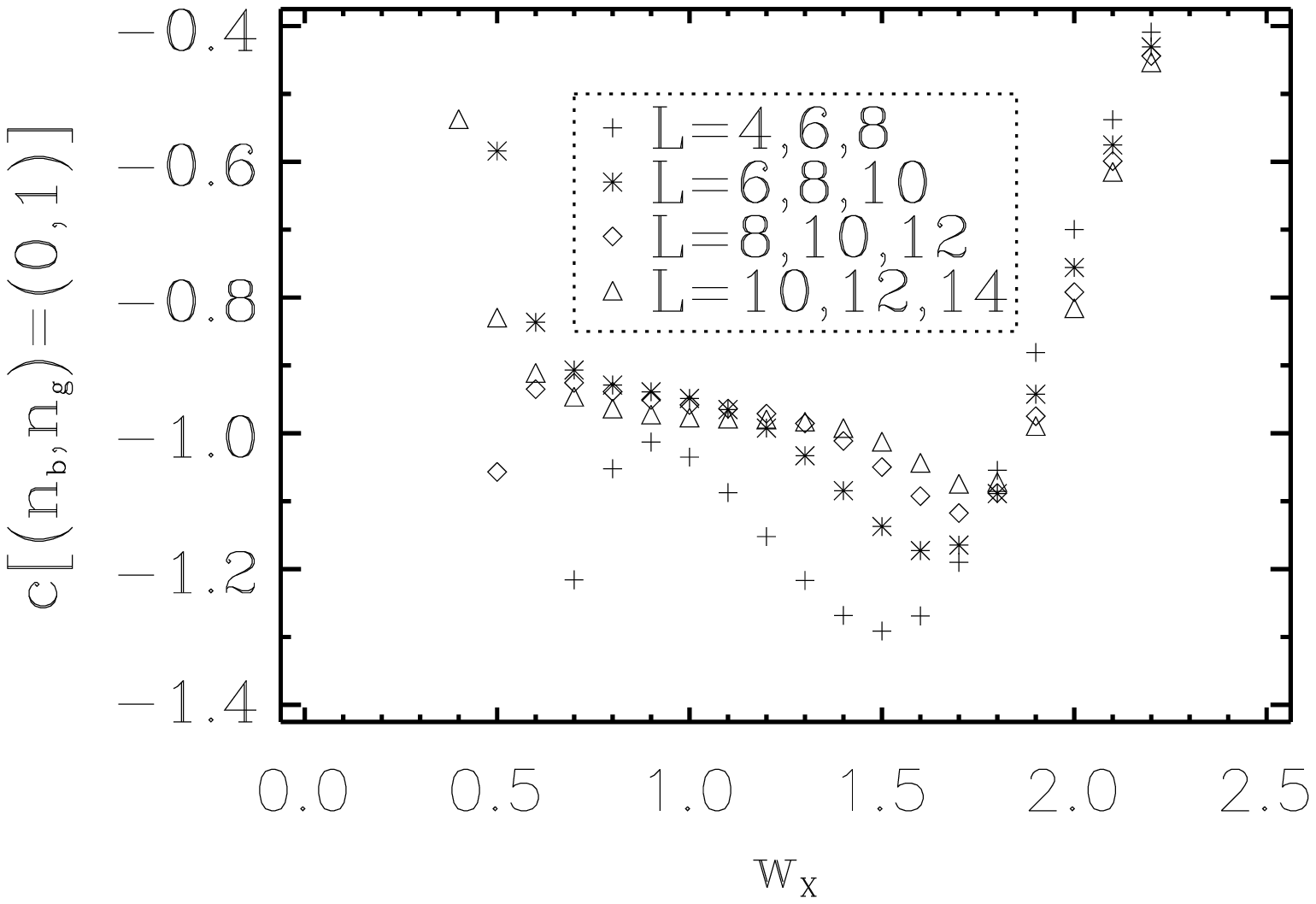}}
 \end{center}
 \protect\caption[3]{\label{fig:cc}Central charge $c$ as a function of the
 bending rigidity $w_{\rm X}$ for the 6V model (left) and the semiflexible loop
 model (right). We show three-point fits for different system sizes, as
 indicated on the legends.}
\end{figure}

In Fig.~\ref{fig:cc} we show the effective central charge as a
function of $w_{\rm X}$, in the cases $(n_{\rb},n_{\rg}) = (1,1)$ and
$(0,1)$. The result $c=1$ for $0 < w_{\rm X} \le 2$ is well established for
the 6V model, but the plot for this case is still useful as it gives
us some guidance as to what finite-size effects to expect. In
particular, note that these become more pronounced when $w_{\rm X}$ is
small, and that the termination of critical behavior at $w_{\rm X}=2$ is
clearly signalled by the finite-size data's levelling off. Another
effect is that while for $w_{\rm X} < 2$ the distance between successive
estimates decreases with system size, for $w_{\rm X} > 2$ we observe this
distance to increase.

Although finite-size effects play a more important role in the
$(n_{\rb},n_{\rg})=(0,1)$ case, the general picture is quite
similar. The figure leaves little doubt that $c=-1$ for $0 < w_{\rm X}
\le w_{\rm X}^{\rm c}$. We also obtain a first rough estimate
$w_{\rm X}^{\rm c} = 1.95 \pm 0.15$, not far away from the 6V-model value.

Here, and elsewhere, we mainly show fits in which the convergence of the
relations Eqs.~(\ref{FSS}) and~(\ref{FSS_2}) has been accelerated through the inclusion of a
non-universal $1/L^4$ correction, as predicted by conformal invariance.

\subsection{String dimensions}

We next turn to the computation of the magnetic-type scaling dimensions
$x_{s_{\rb},s_{\rg}}$ describing the scaling of the operator that inserts
$s_{\rb}$ real strings and $s_{\rg}$ ghost strings. To study
these, the width $L$ of the strip must have the same parity as $s_{\rb}+s_{\rg}$.
For simplicity we shall limit ourselves to the case of even $L$. There
are then two classes of exponents: Those in which $s_{\rb}$ and $s_{\rg}$ are both
even, and those in which they are both odd. The field theory predicts
that the former should stay constant on the critical line, parameterized
by $w_{\rm X}$, while the latter are expected to vary continuously as functions
of $w_{\rm X}$; see Sec.~\ref{string_sec}.

\begin{figure}
\begin{center}
 \leavevmode
 \epsfysize=130pt{\epsffile{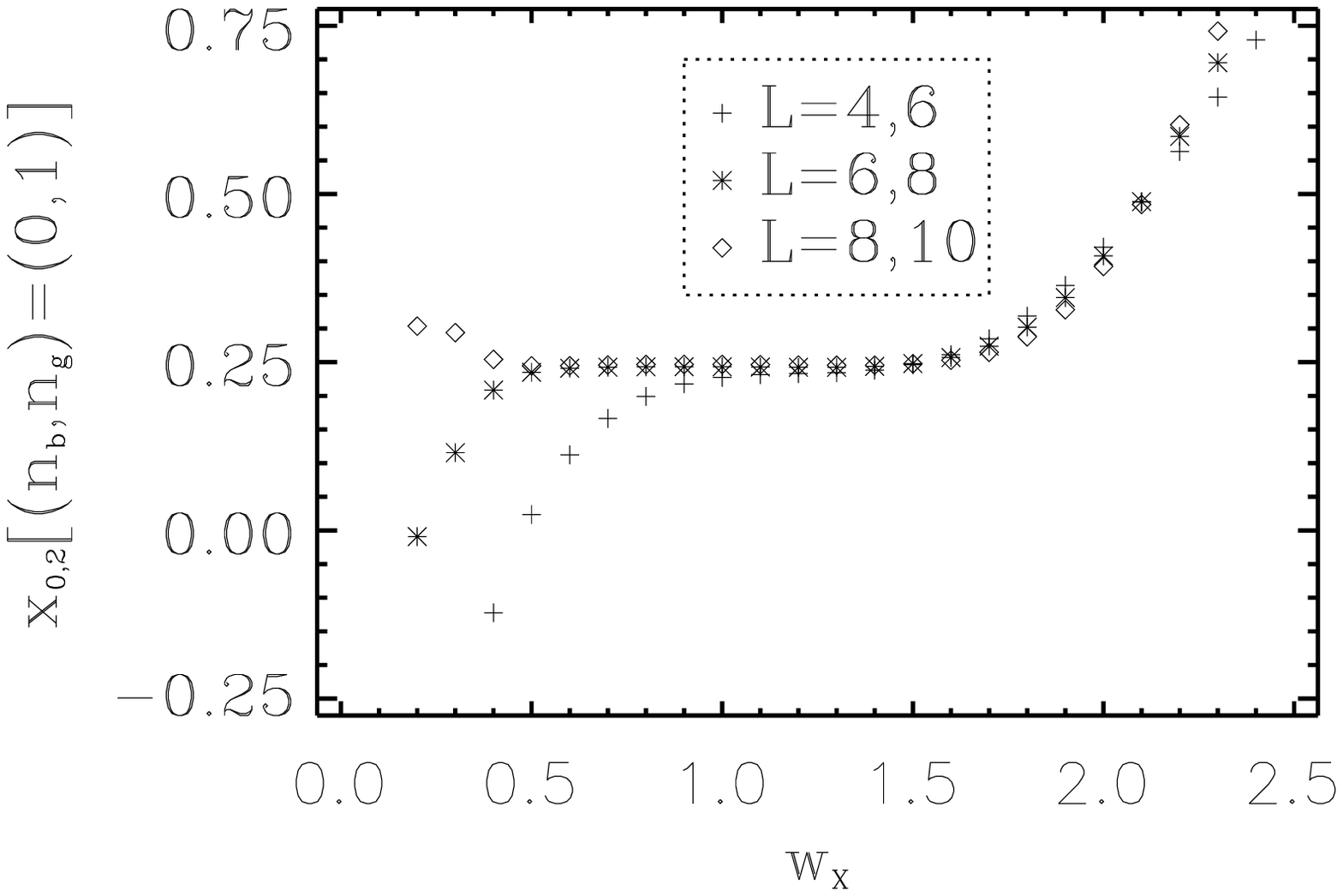}}
 \epsfysize=130pt{\epsffile{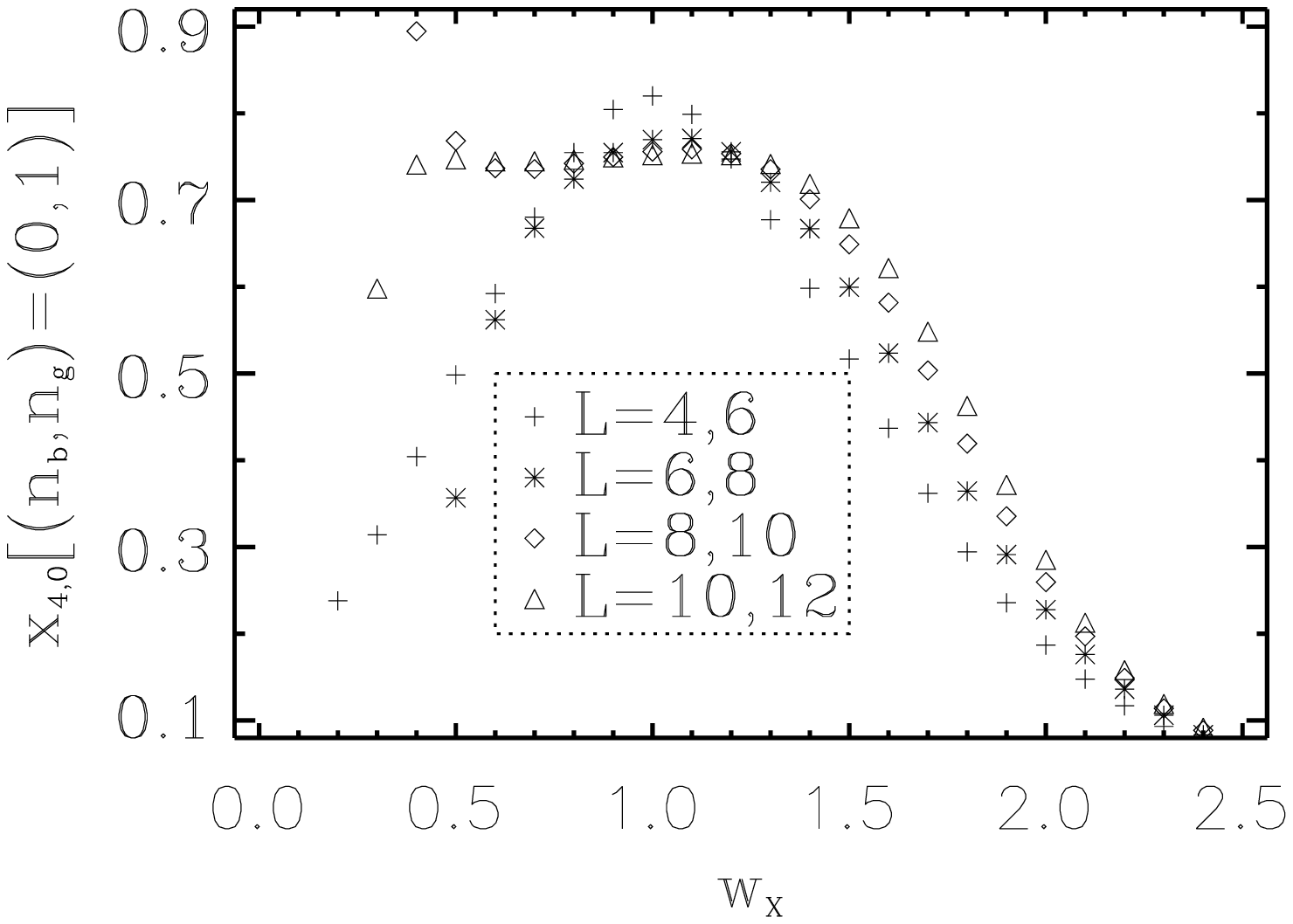}}
 \end{center}
 \protect\caption[3]{\label{fig:xeven}Scaling dimensions $x_{s_{\rb},s_{\rg}}$ with
even $s_{\rb}$ and $s_{\rg}$. The left and right panels show respectively $x_{0,2}$
and $x_{4,0}$. System sizes used in the two-point fits are indicated on the
legends.}
\end{figure}

On Fig.~\ref{fig:xeven} we show two examples of exponents with
$s_{\rb},s_{\rg}$ even, within the SFL model [$(n_{\rb},n_{\rg}) = (0,1)$]
with varying $w_{\rm X}$. They correspond respectively to the insertion of two
ghost strings ($x_{0,2}$) and of four real strings ($x_{4,0}$). From the
figure it should be evident that $x_{0,2}=\frac14$ and $x_{4,0}=\frac34$ are
constant throughout the critical phase. In the latter case the finite-size
variations are quite pronounced, as might have been anticipated given the
higher number of strings. Careful observation of the distance between
subsequent finite-size points however strongly suggests that the variation
will eventually die away.

\begin{figure}
\begin{center}
 \leavevmode
 \epsfysize=130pt{\epsffile{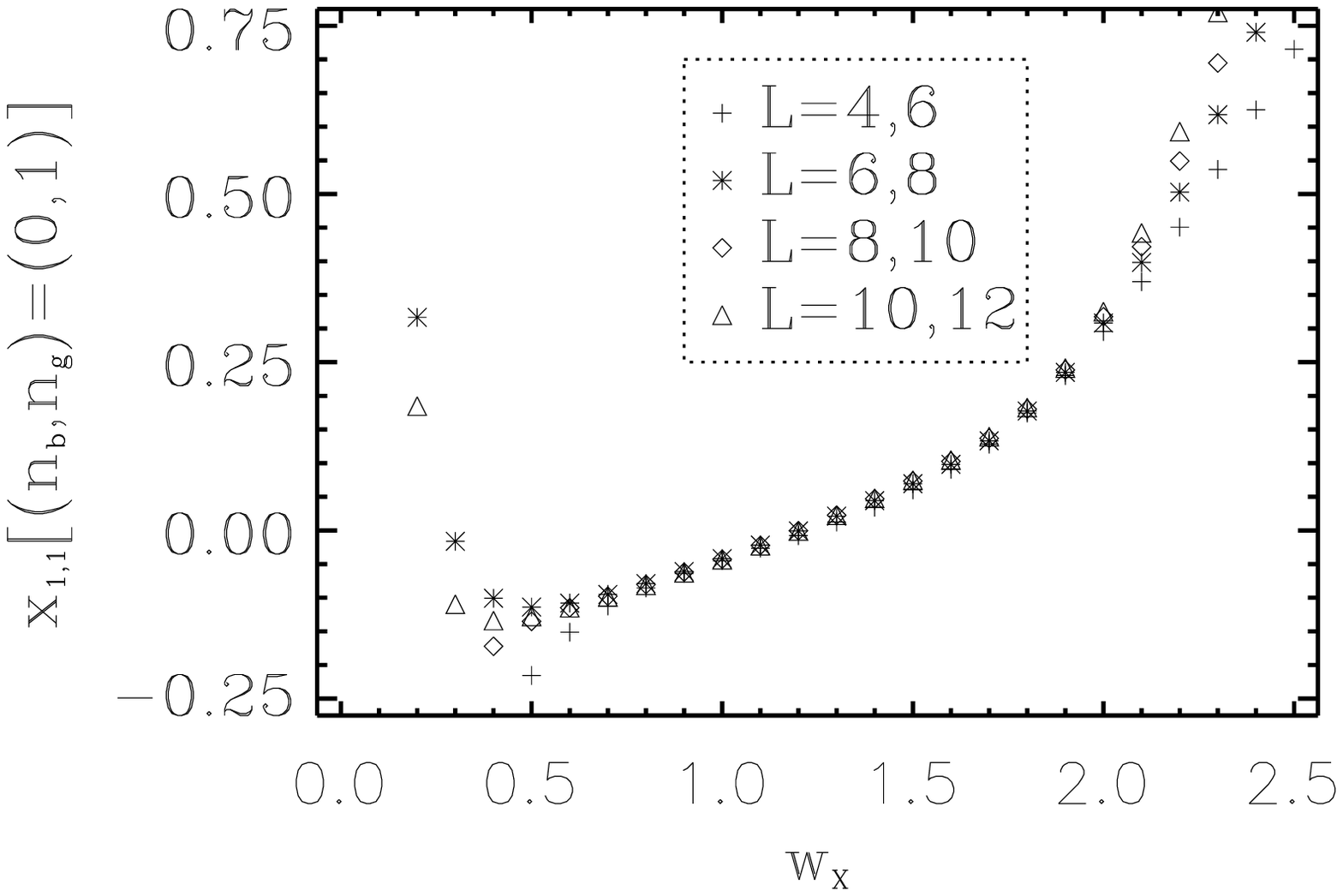}}
 \epsfysize=130pt{\epsffile{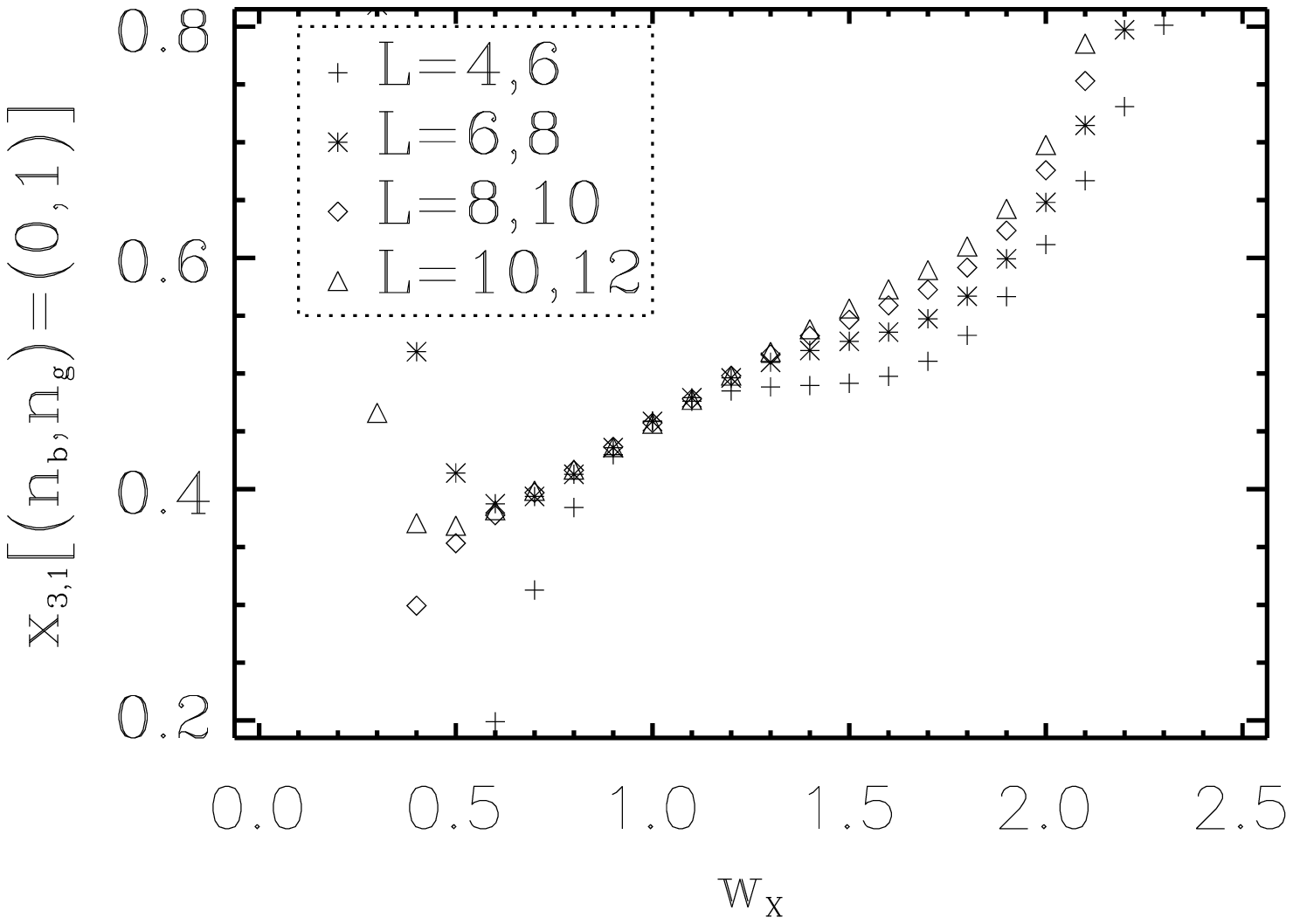}}
 \end{center}
 \protect\caption[3]{\label{fig:xodd}Scaling dimensions $x_{s_{\rb},s_{\rg}}$ with
odd $s_{\rb}$ and $s_{\rg}$. We show $x_{1,1}$ (left) and $x_{3,1}$ (right) for the
SFL model.}
\end{figure}

Examples of exponents with $s_{\rb},s_{\rg}$ odd are given on
Fig.~\ref{fig:xodd}. In both cases, $x_{1,1}$ and $x_{3,1}$, the convergence
to monotonically increasing functions of $w_{\rm X}$ is clearly brought out.
Also note the agreement with the exact results for $w_{\rm X}=1$, which read
respectively $x_{1,1}=-\frac{5}{112} \simeq -0.0446$ and
$x_{3,1}=\frac{51}{112} \simeq 0.455$ \cite{JK_npb}.

\subsection{Thermal scaling dimension}

As described in Appendix~\ref{appA}, the thermal scaling dimension is linked
to the gap between transfer matrix sectors in which there is an even
(respectively an odd) number of flavor crossings in the basis states.
Because of the relation
\be{thermscal}
 K_{22} = \frac{\pi}{4} x_T,
\ee
measuring this gap gives a direct means of accessing the elastic constant
associated with the second height component in the field theory.

According to the field theory, $K_{22}$ is a non-universal function of $w_{\rm X}$,
and once it is known the values of all the other critical exponents follow.
This suggests the following numerical check of the field theoretic scenario:
For several values of $w_{\rm X}$, we measure $x_T$ from the transfer matrix, and
use it to determine $K_{22}$. We then compute the predictions for the various
other scaling dimensions (the $x_{s_{\rb},s_{\rg}}$) from the field theory, \Eq{em_dim},
by use of the numerically determined value of $K_{22}$, and compare them with
values measured directly from the transfer matrices.

\begin{table}
 \begin{center}
 \begin{tabular}{l|lllllll}
   $w_{\rm X}$ & $x_T$    & $x_{1,1}$ & & $x_{3,1}$ & & $x_{1,3}$ & \\
         &          & CFT & Num.  & CFT  & Num. & CFT  & Num. \\ \hline
   0.4   & 0.141(7) & \\
   0.5   & 0.207(5) & \\
   0.6   & 0.275(4) & -0.117 & -0.115 & 0.383 & 0.385 & 0.549 & 0.548 \\
   0.7   & 0.346(3) & -0.100 & -0.100 & 0.400 & 0.402 & 0.567 & 0.565 \\
   0.8   & 0.413(2) & -0.087 & -0.085 & 0.413 & 0.418 & 0.580 & 0.582 \\
   0.9   & 0.4913(7)& -0.065 & -0.066 & 0.435 & 0.436 & 0.601 & 0.601 \\
   1     & $\frac47$& -$\frac{5}{112}$ & & $\frac{51}{112}$ &
                    & $\frac{209}{336}$ &  \\
   1.1   & 0.6525(10)&-0.026 & -0.025 & 0.474 & 0.475 & 0.641 & 0.643 \\
   1.2   & 0.7429(5)& -0.002 & -0.002 & 0.498 & 0.498 & 0.665 & 0.665 \\
   1.3   & 0.8365(3)&  0.022 &  0.022 & 0.522 & 0.521 & 0.688 & 0.688 \\
   1.4   & 0.9374(1)&  0.047 &  0.048 & 0.547 & 0.546 & 0.713 & 0.715 \\
   1.5   & 1.0490(1)&  0.075 &  0.076 & 0.575 & 0.576 & 0.742 & 0.75  \\
   1.6   & 1.1769(7)&  0.107 &  0.11  & 0.607 & 0.62  & 0.774 & 0.80  \\
   1.7   & 1.333(2) &  0.148 &  0.15  & 0.648 & 0.68  & 0.815 & 0.90  \\
   1.8   & 1.541(5) &  0.202 &  0.20  & 0.702 & 0.7   & 0.869 & 0.9   \\
   1.9   & 1.861(8) &  0.29  &  0.3   & 0.79  & 0.8   & 0.95  & 1.0   \\
   $w_{\rm X}^{\rm c}$ & 2&  $\frac{5}{16}$ & & $\frac{13}{16}$ & &
                       $\frac{47}{48}$ & \\
 \end{tabular}
 \end{center}
 \protect\caption[2]{\label{tab:check}Thermal exponent $x_T$ measured for
 varying values of $w_{\rm X}$. The corresponding values of the scaling dimensions
 $x_{1,1}$, $x_{3,1}$ and $x_{1,3}$ are compared with their directly measured
 counterparts.}
\end{table}

The result of this verification is shown in Table~\ref{tab:check}.
The values for $x_T$ are based on transfer matrices for strips up to
size $L=14$, here extrapolated to the limit $L\to\infty$.
The agreement between the CFT predictions and numerics is in general
excellent. Note however, that the precision deteriorates whenever
$w_{\rm X}$ approaches zero or $w_{\rm X}^{\rm c}$, its critical values.

Based on these data we can refine our estimate for the location of the
melting transition:
\be{meltnum}
 w_{\rm X}^{\rm c} = 1.92 \pm 0.02.
\ee

\begin{figure}
\begin{center}
 \leavevmode
 \epsfysize=130pt{\epsffile{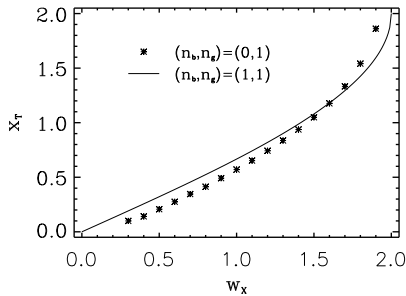}}
 \end{center}
 \protect\caption[3]{\label{fig:coupling}Thermal scaling dimension $x_T$ versus
 bending rigidity $w_{\rm X}$ in the Flory model (symbols), as compared to the
 exact result of the six-vertex model (line).}
\end{figure}

In Fig.~\ref{fig:coupling} we compare our numerical results for the
curve $x_T(w_{\rm X})$ in the SFL case with the exactly known result
of the 6V-model, $x_T=\frac{4}{\pi} \arcsin(w_{\rm X}/2)$
\cite{Baxter}.  Although the functional forms are quite reminiscent,
we have unfortunately not been able to conjecture a convincing exact
expression in the SFL case.

\section{Phase diagram}
\label{sec:pdiagram}

\subsection{Generalized six-vertex model}

Given the one-to-one correspondence between the six vertex configurations
in the FPL${}^2$ model and the six arrow configurations in the six-vertex
model (see Fig.~\ref{fig:6v}), it is natural to define a generalized
six-vertex model in which the standard arrow weights are supplemented by
the non-local loop weights $n_{\rb},n_{\rg}$ of the FPL${}^2$ model.

Until now we have only considered the isotropic case of $a=b$ (see Fig.~\ref{fig:6v}.
Let us briefly recall the effect of taking $a \neq b$ in the six-vertex model \cite{Baxter}.
Define the parameters $w$ and $\mu$ by
\ba{6vparams}
 \Delta &=& \frac{a^2+b^2-w_{\rm X}^2}{2ab} = -\cos \mu, \qquad 0<\mu<\pi \\
 \frac{a}{b} &=& \frac{\exp(i\mu)-\exp(iw)}{\exp(i\mu+iw)-1},
 \qquad -\mu < w < \mu.
\ea
Then, taking $a \neq b$ corresponds to twisting the usual square lattice
into a rhombus, defined by the anisotropy angle \cite{anisotropy}
\be{anisoangle}
 \theta = \frac{\pi}{2} (1+\frac{w}{\mu}).
\ee
All this means is that the central charge and the critical exponents,
when measured in the usual way from a transfer matrix,
get multiplied by a geometrical factor of $\sin(\theta)$.

\begin{figure}
\begin{center}
 \leavevmode
 \epsfysize=150pt{\epsffile{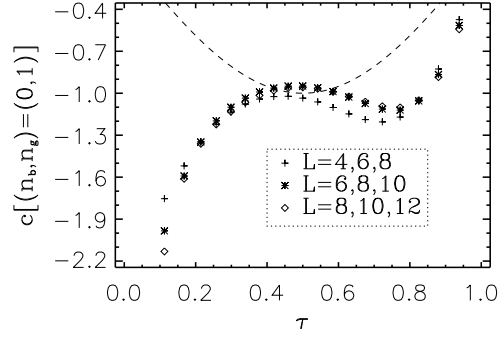}}
 \end{center}
 \protect\caption[3]{\label{fig:aniso}Anisotropy effects in the generalized
 six-vertex model with $(n_{\rm b},n_{\rm g})=(0,1)$.
 The symbols show the effective central charge for various system sizes.
 For comparison, the dashed line shows the function
 $-\sin(\pi \tau)$, which would have been the exact result if the anisotropy
 had had the same effect as in the six-vertex model.}
\end{figure}

In Fig.~\ref{fig:aniso} we plot the effective central charge of the SFL model
with $b=w_{\rm X}=1$ and varying $a$ against the variable
$\tau=\theta/\pi$, defined in terms of the above 6V expressions.
By the word ``effective'' we mean that we do not correct for the lattice
distortion, the effect of which can then be read off from the graph. If
the effect of the anisotropy were the same as in the 6V model, the plot
should just look like the function $-\sin(\pi \tau)$, since the SFL
model has (real) central charge $c=-1$. Clearly, this is
not the case, and so the non-locality of the loop weights has a
non-trivial effect on the anisotropy factor. We leave this as an
interesting open question.

\subsection{Generalized eight-vertex model}
\label{sec:8v}

\begin{figure}
 \begin{center}
 \leavevmode
 \epsfysize=70pt{\epsffile{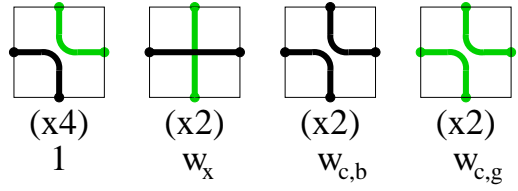}}
 \end{center}
 \protect\caption[3]{\label{fig:8v}Vertices defining the generalized
 eight-vertex model, along with their corresponding multiplicities and local weights.}
\end{figure}

It is also of interest to consider the loop generalization of the
eight-vertex model. In terms of the loops there are two different ways of
resolving the vertices that act as sources or sinks of the eight-vertex
arrows, and so we are led to consider the ten-vertex model defined by
Fig.~\ref{fig:8v}. In addition to the local weights which are shown on
the figure, we assign the usual non-local loop weights $n_{\rb}$ and $n_{\rg}$.

For simplicity, we shall disregard the effects of anisotropy, and thus
only two types of local weights are of interest. The first is the weight
$w_{\rm X}$ of having the two loop flavors cross, same as in the SFL model.
The second is a contact interaction $w_{\rm c}$, assigned to the vertices where
two loop segments of the same flavor touch one another. One may consider
letting it depend on the flavor index, but in order to stay close to
the definition of the conventional eight-vertex model we shall here take
the contact interaction to be flavor independent.

The motivation for the contact interaction is to be able to exclude the
loops of a given flavor from any number of  lattice vertices. As this
violates the compactness constraint, we expect the conclusions of our
earlier paper on the transition from the compact to the dense phase
\cite{JK_jsp} to apply: A non-zero value of $w_{\rm c}$ should induce a flow
towards a phase where the two loop flavors decouple, and the critical
properties are just those of two non-interacting O($n$) models (with
$n=n_{\rb}$ and $n_{\rg}$ respectively) in the low-temperature (dense) phase.

\begin{figure}
 \begin{center}
 \leavevmode
 \epsfysize=150pt{\epsffile{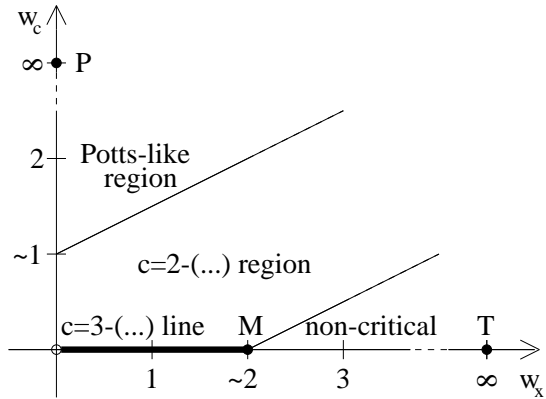}}
 \end{center}
 \protect\caption[3]{\label{fig:8v_pd}Proposed phase diagram of the
 generalized eight-vertex model.}
\end{figure}

A detailed numerical study of the behavior of the effective central charge
in the parameter space $(w_{\rm X},w_{\rm c})$ has led us to suggest that the phase diagram
of the generalized eight-vertex model is as shown on Fig.~\ref{fig:8v_pd}.

For $w_{\rm c}=0$, the model reduces to the SFL model, and so below the melting
point M (i.e., for $w_{\rm X} < w_{\rm X}^{\rm c}$ with, very roughly,
$w_{\rm X}^{\rm c} \approx 2$) we have a line of critical points
along which critical exponents that depend on  the second height
component vary continuously, while the central charge
\be{ccSFL}
 c(n_{\rb},n_{\rg}) = 3 - \frac{6e_{\rm b}^2}{1-e_{\rm b}} -
 \frac{6e_{\rm g}^2}{1-e_{\rm g}}
\ee
is constant. The end-point of the SFL line, with $w_{\rm X}=\infty$, is a
trivial attractive fixed point T, favoring configurations in which all loops
go straight in the bulk (they are necessarily reflected at the boundaries
enjoying free boundary conditions). The point T is believed to govern the
theories to the right of M (i.e., with $w_{\rm X} > w_{\rm X}^{\rm c}$),
including a portion of the phase diagram with non-zero but small $w_{\rm c}$
(see Fig.~\ref{fig:8v_pd}).

Moving away from the critical line of the SFL model, towards positive values
of the contact interaction, we observe numerically that the central charge
drops abruptly by one unit, and stays constant as a function of $w_{\rm c}$ up
to some finite critical value $w_{\rm c}^{\rm c}$ that depends on $w_{\rm X}$.
This is the dense phase of the DPL${}^2$ model \cite{JK_jsp} with central
charge
\be{ccDPL}
 c(n_{\rb},n_{\rg}) = \left( 1 - \frac{6e_{\rm b}^2}{1-e_{\rm b}} \right) +
              \left( 1 - \frac{6e_{\rm g}^2}{1-e_{\rm g}} \right).
\ee
Here, the two loop flavors decouple, and critical exponents are just the sum
of the critical exponents for two non-interacting O($n$) models (with
$n=n_{\rb}$ and $n_{\rg}$) in the dense phase. We have verified numerically
this prediction for the exponent $x_{1,1}$ for a number of different loop
fugacities. We have also observed numerically that the critical exponents
do not depend on $w_{\rm X}$ throughout the dense phase. This confirms the
expectations that in non-compact phases the only effect of the bending
rigidity is to renormalize the persistence length of the polymer, as
already discussed in the introduction.

Finally, for $w_{\rm c}$ large enough, the numerically evaluated central
charges suggest that the models flow into an attractive fixed point P situated
at $(w_{\rm X},w_{\rm c}) = (0,\infty)$. Here, only contact-type vertices are
allowed, and since the different loop flavors can no longer coexist, the
partition function at P becomes simply a sum, $Z = Z_{\rm b} + Z_{\rm g}$,
where $Z_k$ involves only contact vertices of flavor $k$ (with $k={\rm b},{\rm
g}$). But clearly $Z_k$ is just the loop-model representation \cite{Baxter} of
a self-dual Potts model with $q_k = (n_k)^2$ states. It is intuitively clear
(and explicitly brought out by the exact solution \cite{Baxter}) that the free
energy of the $q$-state Potts model is an increasing function of $q$.
Therefore, the sum $Z=Z_{\rm b}+Z_{\rm g}$ will be dominated by the term with
the largest value of $q$. Thus, the point P has central charge
\be{ccPotts}
 c(n_{\rb},n_{\rg}) = \mbox{max}\left( 1 - \frac{6e_{\rm b}^2}{1-e_{\rm b}},
                               1 - \frac{6e_{\rm g}^2}{1-e_{\rm g}} \right),
\ee
and, by the usual identification of the critical Potts model with the dense
phase of the O($n=\sqrt{q}$) model, the critical exponents are simply those of
a single O(max($n_{\rb},n_{\rg}$)) model in the dense phase.

We would expect that only this large-$w_{\rm c}$ portion of the phase
diagram gets modified by letting the contact interaction be flavor
dependent. Let us recall that in the conventional O($n$) model
\cite{nienhuis82} with a finite {\em positive} vacancy fugacity
$w_{\rm c}$ the critical behavior of the loops is described by either
of {\em two} critical branches. The first branch, known as the {\em
dense} branch \cite{nienhuis82,DupSal87}, is attractive in $w_{\rm
c}$ and as such controls the entire domain of low $w_{\rm c}$. Its
central charge is the one referred to above:
\be{c-dense}
 c=1-6 \tilde{e}^2/(1-\tilde{e})
\ee
in the usual parameterization $n=2 \cos(\pi \tilde{e})$. The second
branch, known as the {\em dilute} branch \cite{nienhuis82,saleur86b},
is repulsive in $w_{\rm c}$
and as such requires $w_{\rm c}$ to be tuned to a particular
$n$-dependent critical value. In other words, the fugacity of a
vacancy can tune the O($n$) model to its critical point. The central
charge of the dilute phase is
\be{c-dilute}
 c=1-6 \tilde{e}^2/(1+\tilde{e}) \ ,
\ee
using the same parameterization as above.

In particular, in the DPL${}^2$ model \cite{JK_jsp} the two loop flavors
act as decoupled O($n$) models, and depending on the fugacities of the
two flavors of vacancies each of the models can reside in either the
dense or the dilute phase, giving a total of four different phases.
We expect this conclusion to hold true in the generalized eight-vertex
model (i.e., with an added bending rigidity $w_{\rm X}$). Note that only
when $n_{\rm b}=n_{\rm g}$ can we simultaneously take the two
O($n$) models to their critical point
by tuning a vacancy fugacity $w_{\rm c}$ which is common for the two
loop flavors. In the general case, when $n_{\rm b} \neq n_{\rm g}$
we would need two distinct parameters,
$w_{\rm c,b}$ and $w_{\rm c,g}$, as indicated on Fig.~\ref{fig:8v}.
Presumably this would lead to a richer phase diagram, with critical
lines corresponding to dense-dilute, dilute-dense and dilute-dilute
behavior of the two O($n$) models.

Let us return to the phase diagram shown in Fig.~\ref{fig:8v_pd}. In the
special case of the eight-vertex model, the two oblique transition lines shown on
Fig.~\ref{fig:8v_pd} are known to be of slope $1/2$
\cite{Baxter}. Actually, they are just images of the line OM under certain
exact symmetries of the eight-vertex model \cite{Baxter}. Thus, they have
again $c=1$, whereas the ``bulk'' of the phase diagram is non-critical.

These two features can be accounted for within the framework of the
generalized eight-vertex model with $(n_{\rb},n_{\rg})=(1,1)$.
First, note that our field theory predicts that the region on
Fig.~\ref{fig:8v_pd} which is limited by the two oblique lines and the
coordinate axes is actually {\em critical} with central charge $c=0$ (dense
phase); this is obtained by setting $e_{\rm b}=e_{\rm g}=1/3$ in
\Eq{ccDPL}. This is not in contradiction with the exact result
\cite{Baxter} that this same region is non-critical within the
eight-vertex model. Namely, the generalized eight-vertex model is
embedded in a much larger Hilbert space. More precisely, our statement
is that the first and third height components possess critical
fluctuations and constitute a $c=0$ theory, even though the second
height component is massive.
This scenario is brought out very clearly by the numerics, as we observe
the leading transfer matrix eigenvalues in the sectors determining the free energy
and $x_T$ to {\em coincide} within the concerned region. Thus, $x_T$ and
$K_{22}$ vanish identically, cf.~Eq.~(\ref{thermscal}).

Second, the field theory also accounts for the fact that, within the
8V model, the two oblique lines are critical with $c=1$. Namely, we
claim that they simply correspond to dilute-phase behavior within the
generalized 8V model. More precisely, since $n_{\rm b}=n_{\rm g}$
the two decoupled O($n$) models must be driven to their critical points
simultaneously by tuning the common parameter
$w_{\rm c,b}=w_{\rm c,g} \equiv w_{\rm c}$.
Setting $\tilde{e}=1/3$ in \Eq{c-dilute} gives a contribution of
$c=1/2$ for each of the models, whence $c_{\rm total}=1/2+1/2=1$ as
expected.

Finally, in the 8V model, the part of the $w_{\rm c}$ axis with
$0<w_{\rm c}<1$ constitutes a further image of the line OM under an
exact symmetry. We believe this to be ``accidental'' in the sense that
we have seen no sign of a finite interval of the $w_{\rm c}$ being
critical within the generalized 8V model with other values of the fugacities.

Taking a common contact parameter $w_{\rm c,b}=w_{\rm c,g}$ for the
generalized 8V model with $n_{\rm b} \neq n_{\rm g}$ destroys the
criticality of the two oblique lines of Fig.~\ref{fig:8v_pd}. They still
act as transition lines in the sense that they separate the basins of
attraction of the dense phase and the points P and T respectively.
However, the transition is now expected to be a first order one.
This is confirmed by our numerical results for the
$(n_{\rm b},n_{\rm g})=(0,1)$ case which show that the effective central
charge develops violent finite-size effects upon approach of the
transition lines.
Further support for this scenario is furnished by Monte Carlo simulations
\cite{yoon} where a finite concentration of empty sites was shown to lead
to a first order transition.

The oblique lines in Fig.~\ref{fig:8v_pd} are expected to move away
from their exactly known 8V positions when we vary the loop fugacities
away from the trivial values ($n_{\rm b}=n_{\rm g}=1$).  Some evidence
for this is already available from our determination of the melting
point M in the Flory case; see \Eq{meltnum}. In general, we have been able
to numerically determine the position of the uppermost line from the
transfer matrix spectra. Recall from the discussion near \Eq{thermscal} that
the coupling $K_{22}$ can be linked to the gap between the leading eigenvalues
in two topologically characterized transfer matrix sectors. By scanning
through $w_{\rm c}$ at fixed $w_{\rm X}$ we have observed (at least in the
Flory case) that these two eigenvalues become degenerate as soon as
$w_{\rm c}$ moves away from zero (even at a value as small as
$w_{\rm c} \simeq 10^{-6}$). This degeneracy eventually disappears when
there is a level crossing in the groundstate sector of the transfer matrix,
at some finite $w_{\rm c}$.

We have measured the position of this level crossing as a function of
system size and extrapolated it to the thermodynamic limit. To test the
reliability of the method, we have first applied it to the
$(n_{\rm b},n_{\rm g})=(1,1)$ case. Our final estimate
$w_{\rm c}=1.52 \pm 0.02$ at fixed $w_{\rm X}=1$ is
in good agreement with the exact result $w_{\rm c}=3/2$ \cite{Baxter}.
The same method applied to the Flory case, $(n_{\rm b},n_{\rm g})=(0,1)$,
yields $w_{\rm c}=1.4294 \pm 0.0005$ at $w_{\rm X}=1$
and $w_{\rm c}=1.958 \pm 0.005$ at $w_{\rm X}=2$. These values are
clearly different from those predicted by the 8V model.

In conclusion, we believe that it would be most interesting to study
the generalized eight-vertex model in more detail, using the exact
techniques of integrable systems. In particular, it is conceivable
that the present treatment misses some subtle exceptional points in
the phase diagram.

\section{Discussion}

The semiflexible loop model was defined as a generalization of the two-flavor
fully packed loop model on the square lattice, by introducing a vertex weight
associated with vertices at which the loop does not undergo a 90${}^\circ$
turn. We have proposed an effective field theory of the semiflexible loop model
based on its height representation. This leads to exact results for the Flory
model of polymer melting in two dimensions. Furthermore we have shown that the
loop model provides a generalization of the eight-vertex model with an
interesting phase diagram. Here we comment further on these two main results.

\subsection{Scaling of semiflexible compact polymers}

Polymers configurations are random and as such they are described by
probability distributions.  Their critical nature, in the long chain
limit, is revealed by the fact that these distributions have scaling
forms characterized by universal exponents. The simplest distribution
is the probability $p(r,l)$ that the end-to-end distance equals $r$
for a polymer of contour length $l$. In the scaling limit, when $r$ is
much greater than the lattice spacing and much less than the radius of
gyration of the polymer, we have \cite{degennes}
\be{pol_sc}
 p(r,l) = r^{\theta} f(r/l^\nu) \  .
\ee
Here $f$ is a scaling function, $\nu$ is the `` swelling exponent'',
and $\theta$ the ``cyclization exponent''.

For semiflexible compact polymers, which correspond to the Flory model
with $w_{\rm X}\le w^{\rm c}_{\rm X}$, the swelling exponent is $\nu =
1/2$. This is an exact result which simply follows from the fact that
compact polymers are space filling, regardless of $w_{\rm
X}$. Furthermore, the swelling exponent can be related to the string
dimension $x_{2,0}$ through the scaling law $\nu =
(2-x_{2,0})^{-1}$ \cite{JK_npb}. Then replacing $e_{\rb}=1/2$ and $e_{\rg}=1/3$ in
\Eq{x_even} gives $x_{2,0}=0$ and $\nu=1/2$, for all values of $w_{\rm
X}$. This calculation then serves as a non-trivial check on the field
theory.

The cyclization exponent is related to the scaling dimension
associated with one real and one ghost loop segment: $\theta = - 2
x_{1,1}$ \cite{JK_npb}. From \Eq{x_odd} it follows that $\theta$ will vary
continuously as the polymer is made stiffer by increasing $w_{\rm X}$.
At the melting transition the exact result for $\theta$ follows from
the computed value of the critical elastic constant, \Eq{stiff22_c},
\be{theta_crit}
\theta^{\rm c} = -\frac{5}{8} \ .
\ee
The negative value implies that at the transition (and slightly below
it) the two ends of the polymer feel an effective {\em
attraction}. This is surprising as the naive expectation is that the two
ends of a polymer will feel an effective repulsion due to the
self-avoiding constraint. For stiff compact polymers this naive
expectation is not met. Whether this will persist in three dimensions
is an interesting open question.

\subsection{Generalized eight-vertex model}

The generalized 8V model gives a quite detailed modelization of two-dimensional
lattice polymers. It possesses the following features:
\begin{itemize}
 \item Steric constraints (self-avoidance and connectedness
       of the polymer chains) are modelled exactly;
 \item Possibility of introducing polydispersity, by taking $n_{\rm b}$
       away from zero;
 \item A bending rigidity parameter $w_{\rm X}$ allows to control the
       transition between a melt and a crystalline phase;
 \item A contact interaction parameter $w_{\rm c}$ (or alternatively a
       fugacity of a vacancy) controls the transitions between compact,
       dense, dilute, Potts-like and non-critical phases;
 \item Possibility of introducing non-local interactions (although of a
       peculiar form), by taking $n_{\rm g}$ away from one.
\end{itemize}

The phase diagram of a somewhat similar model was studied in the Bethe
approximation by Lise, Maritan, and Pelizzola \cite{lmp98}. However, in the
compact limit the results of these authors are equivalent to Flory's
mean-field treatment, as they do not take into account the non-local features
of the polymers. We have here treated the excluded-volume effects in an
exact manner. On the other hand, the model of Ref.~\cite{lmp98} includes
an additional feature:
\begin{itemize}
 \item A contact interaction between non-consecutive monomers that are
       nearest neighbors on the lattice allows to drive the model to
       tricriticality, i.e., to access the theta-point physics.
\end{itemize}
This interaction is not present in the generalized 8V-model. If we were to
include it, we would need the contact interaction to be flavor dependent (the
authors of Ref.~\cite{lmp98} do not consider what we refer to as ghost loops).
However, we do not believe that anything new can be learnt from such a
generalization. First, the contact interaction is redundant in the compact
phase ($w_{\rm c}=0$), as the number of contacts is constant (actually
maximal) in any fully packed configuration. Second, in the non-compact phases
($w_{\rm c} \neq 0$) our field theory predicts a decoupling into two
independent O($n$) models. One would expect the flavor-dependent contact
interactions to act independently on the two decoupled models, and the problem
essentially reduces to that of the theta-point physics of a standard O($n$)
model \cite{theta}.

We leave it as an interesting question whether the generalized 8V model
can be tackled using the methods of integrable systems. {}From
Fig.~\ref{fig:8v} it can obviously be formulated as a forty-vertex model
(taking into account the loop orientations) with complex vertex weights.
To our knowledge, such a model has not been studied previously. If one
could solve it, it would be particularly interesting to work out the exact
expression of the coupling constant $K_{22}(w_{\rm X})$ as a function of
the loop fugacities $n_{\rm b}$ and $n_{\rm g}$. Also, it is conceivable
that our field theoretical approach has missed some exceptional critical
points in the phase diagram.

\subsection{Order of the melting transition}

In this paper we have established that the order of the melting transition
within the Flory model is second order, as first suggested by Saleur \cite{saleur}.
We have also explained how the introduction of a finite density of vacancies
may lead to a first order transition, as observed in Monte Carlo simulations
\cite{yoon}. This combined scenario settles a long controversy in the literature \cite{menon}.

\appendix

\section{Construction of the transfer matrices}
\label{appA}

The transfer matrix construction of Ref.~\cite{JK_npb} relied on an explicit
bijection between the set of allowed connectivity states ${\cal C}$ and the
set of integers $Z_{|{\cal C}|} = \{1,2,3,\ldots,|{\cal C}|\}$. However, in
many cases it is difficult to furnish an {\em a priori} characterization of
the set of allowed basis states and its cardinality. Moreover, some of the
states utilized in \cite{JK_npb} were found to carry zero weight in the
leading eigenvectors of the corresponding sectors of the transfer matrix, and
so one should think that it would be possible to eliminate them from the
outset.

To remedy this situation it is preferable to use another approach. Without
prior knowledge of the state space, the latter is explicitly generated by
acting with the transfer matrix ${\cal T}$ on a reference state
$|v_0\rangle$ which is known to belong to the image of ${\cal T}$ in the
concerned sector. In this way, a certain number of image states is generated,
which can be inserted in an appropriate data structure using hashing
techniques \cite{hashing}. One then acts with ${\cal T}$ on these states,
generating a new list of states, and continues in this way until no new
states are generated. The resulting list is the complete state space of
${\cal T}$ in the concerned sector.

\begin{figure}
 \begin{center}
 \leavevmode
 \epsfysize=50pt{\epsffile{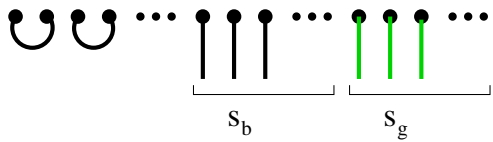}}
 \end{center}
 \protect\caption[3]{\label{fig:refstate}Reference state used for generating
 the sector with $s_k$ flavor-$k$ strings ($k={\rm b},{\rm g}$).}
\end{figure}

It remains to find an appropriate reference state $|v_0\rangle$ for each
physically interesting sector of ${\cal T}$. For the sector
$(s_{\rb},s_{\rg})$ in which $s_k$ flavor-$k$ strings ($k={\rm b},{\rm g}$)
span the length of the cylinder generated upon action of ${\cal T}$, the
reference state can be chosen as shown in Fig.~\ref{fig:refstate}. This state
simply consists of $(L-s_{\rb}-s_{\rg})/2$ real arches followed by
$s_{\rb}$ real strings and $s_{\rg}$ ghost strings. $L$ must of course
have the same parity as $s_{\rb}+s_{\rg}$.

Choosing the sector corresponding to the thermal scaling dimension is a little
less obvious. A useful observation is made by exploiting the correspondence
with states of the six-vertex model, as depicted in Fig.~\ref{fig:6v}. In a
given row, let $N_k^{\rm even}$ and $N_k^{\rm odd}$ be the number of
flavor-$k$ loop segments residing on even and odd vertical edges,
respectively. Then define
\be{flux}
 Q = \left( N_{\rb}^{\rm even}-N_{\rb}^{\rm odd} \right) -
     \left( N_{\rg}^{\rm even}-N_{\rg}^{\rm odd} \right).
\ee
By inspection of Fig.~\ref{fig:6v} it is seen that $Q$ is nothing but the
vertical flux of arrows within a given row. By the ice rule, $Q$ is a
conserved quantity and can thus be used to label a sector of ${\cal T}$.

\begin{figure}
 \begin{center}
 \leavevmode
 \epsfysize=20pt{\epsffile{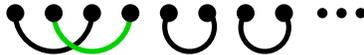}}
 \end{center}
 \protect\caption[3]{\label{fig:excstate}Reference state used for generating
 the thermal sector.}
\end{figure}

The reference state of Fig.~\ref{fig:refstate} with $(s_{\rb},s_{\rg})=(0,0)$
is seen to have $Q=0$. The first excited state with no strings has
$Q=\pm 4$ and is depicted on Fig.~\ref{fig:excstate}. Its first four
sites are occupied by mutually penetrating real and ghost arches,
followed by $(L-4)/2$ simple real arches. In general,
for any given $L$, states with $Q=\pm 4q$ exist for
$q=0,1,\ldots,\lfloor L/4 \rfloor$. The number of states in the $k$'th
sector is just
\be{states1}
 \sum_{n=q}^{L/2-q} {L/2 \choose n+q} {L/2 \choose n-q} C_{L/2-n} C_n,
\ee
where $C_n = \frac{(2n)!}{n! (n+1)!}$ are the Catalan numbers.
Using a sum rule on the binomial coefficients, it is easily seen that the
total number of states without strings, summed over the sector index $q$,
reads simply
\be{states2}
 \sum_{n=0}^{L/2} {L \choose 2n} C_{L/2-n} C_n.
\ee
This is nothing but the dimension of the state space used in \cite{JK_npb}.

{}From entropic reasons it is fairly obvious that the free energy belongs to
the sector $q=0$. We are now going to argue that the thermal scaling exponent
is linked to the gap between the first eigenvalue in the $q=0$ and $q=1$
sectors, cf.~\Eq{FSS_2}. The first reason is that, by construction,
the $q=1$ constraint acts as an excitation within the full state space (with
all values of $q$ included, as in \Eq{states2}), and hence should
correspond to a subdominant eigenvalue within that space. Indeed, it is observed
numerically that the second eigenvalue obtained from the transfer matrix of
Ref.~\cite{JK_npb} coincides with the leading eigenvalue of the $q=1$ sector,
obtained by using the techniques outlined above.

As a second argument, note that in the language of the SFL model height
mapping, encircling the first four sites of Fig.~\ref{fig:excstate}
yields a height dislocation of ${\bf A}-{\bf C}+{\bf B}-{\bf D}$. By the
four-coloring rule, ${\bf A}+{\bf B}+{\bf C}+{\bf D}={\bf 0}$, this is the
same as
\be{thermdef}
 {\bf m}_T = 2({\bf A}+{\bf B}) = -2({\bf C}+{\bf D}).
\ee
But the latter is {\em also} the height defect associated with a defect vertex
$({\bf C},{\bf D},{\bf C},{\bf D})$ that corresponds to excluding the real
loops from that vertex, which is exactly a thermal-type excitation (and to wit
the one that is used for computing the critical exponent $x_T$ within the
field theory).

In the field theory, one might compute the exponent corresponding to the
insertion of magnetic defects $\pm q' {\bf m}_T$ at either end of the cylinder.
In the transfer matrix, these should simply be linked to the gap between the
sectors $q=0$ and $q=q'$.

\begin{table}
 \begin{center}
 \begin{tabular}{r|rrrrrrr}
   $L$ & $(0,0)$ & Thermal & $(2,0)$ & $(1,1)$ & $(4,0)$ & $(3,1)$ & $(2,2)$
   \\ \hline
     2 &       2 &     --- &       1 &       1 &     --- &     --- &     --- \\
     4 &       8 &       1 &       8 &      12 &       1 &       2 &       2 \\
     6 &      46 &      12 &      69 &     141 &      15 &      42 &      72 \\
     8 &     332 &     124 &     664 &    1720 &     196 &     684 &    1056 \\
    10 &    2784 &    1280 &    6960 &   21760 &    2520 &   10320 &   14800 \\
    12 &   25888 &   13605 &   77664 &  283584 &   32565 &  151500 &  205920 \\
    14 &  259382 &  149604 &  907837 &         &  425019 &         &         \\
 \end{tabular}
 \end{center}
 \protect\caption[2]{\label{tab:size}Sizes of various sectors of the SFL model
 transfer matrix defined on a cylinder of width $L$ with periodic boundary
 conditions. The symbol $(s_{\rb},s_{\rg})$ labels the sector in which $s_k$
 flavor-$k$ strings ($k={\rm b},{\rm g}$) run along the length of the cylinder.
 The sectors $(0,0)$ and `Thermal' have no strings, but the parity of the
 number of flavor-crossings in the basis states is fixed to be even and odd,
 respectively.}
\end{table}

In Table~\ref{tab:size} we show the sizes of the various transfer matrices
used in this work. The columns labelled $(0,0)$ and `Thermal' correspond
to the expressions (\ref{states1}) with $q=0$ and $q=1$, respectively.
For the other columns, similar expressions may be worked out along the lines
of \cite{JK_npb}.

Finally, let us remark that the computations for the generalized eight-vertex
model introduced in Section~\ref{sec:8v} are produced from the same reference
states, but slightly generalizing the transfer matrix to accommodate the
contact-type vertices shown on Fig.~\ref{fig:8v}.

\acknowledgements We are grateful to Ken Dill for introducing us to the Flory model
of polymer melting. JK would further like to thank the KITP in Santa Barbara for
hospitality, where this work was initiated. The research of JK is supported by the
NSF under grant number DMR-9984471. JK is a Cottrell Scholar of Research Corporation.

\end{document}